\documentclass{LMCS}

\usepackage{ifthen}
\newboolean{long}
\setboolean{long}{false}
\usepackage{url}
\usepackage{enumerate}
\usepackage{hyperref}

\usepackage{color}
\usepackage{amssymb}

\usepackage{amsmath}
\usepackage{amsfonts}
\usepackage{proof}
\usepackage{611}

\theoremstyle{plain}
\newtheorem{corollary}[thm]{Corollary}
\newtheorem{lemma}[thm]{Lemma}
\theoremstyle{definition}
\newtheorem{definition}[thm]{Definition}

\newcommand\iffl{\ensuremath{\leftrightarrow}}

\renewcommand\int{\mathbf{Z}}

\newcommand{\ignore}[1]{}
\newcommand{\todo}[1]{}
\newcommand{\ov}[1]{\ensuremath{\vec{#1}}}
\newcommand{\nat}{\ensuremath{\mathbb{N}}}

\newcommand\reals{\ensuremath{\Vdash}}
\newcommand{\rrho}{\reals_\rho}
\newcommand{\p}{\proves}

\newcommand{\g}{\Gamma}
\newcommand{\gp}{\Gamma \proves}

\newcommand{\og}{\ov{\g}}

\newcommand{\pl}[1]{\ensuremath{\mathrm{#1}}}
\newcommand{\FST}{\pl{fst}}
\newcommand{\SND}{\pl{snd}}
\newcommand{\LET}{\pl{let}}
\newcommand{\CASE}{\pl{case}}
\newcommand{\MAGIC}{\pl{magic}}
\newcommand{\INL}{\pl{inl}}
\newcommand{\INR}{\pl{inr}}
\newcommand{\IN}{\pl{in}}
\newcommand{\IND}{\pl{ind}}
\newcommand{\sr}[1]{\ensuremath{\SB{#1}_\rho}}
\newcommand{\izfi}{IZF${}^{-}_{R \omega}$}
\newcommand{\izfio}{IZF${}_{R \omega}$}
\newcommand{\phiinaca}{\ensuremath{\phi^i_1}}
\newcommand{\phiinacb}{\ensuremath{\phi^i_2}}
\newcommand{\eqo}{=}
\newcommand{\vis}[1]{\ensuremath{U_{#1}}}
\newcommand{\rphi}{\phi}
\newcommand{\sphi}{\phi}
\newcommand{\iphi}{\phi}
\renewcommand{\ini}{\in_I}
\newcommand{\ino}{\in}
\newcommand{\rin}[1]{\ensuremath{#1^+}}

\newcommand{\izfc}{IZF${}_C$}
\newcommand{\izfr}{IZF${}_R$}
\newcommand{\izfro}{IZF${}_{R0}$}
\newcommand{\inac}[1]{\ensuremath{\Gamma_{#1}}}
\newcommand{\vinac}[1]{\ensuremath{V^\lambda_{\inac{#1}}}}
\newcommand{\lrk}{\ensuremath{\lambda rk}}

\newcommand{\li}{\lambda Z_\omega}
\newcommand{\la}{\lambda \overline{Z_\omega}}
\newcommand{\lat}{\Lambda_{\overline{Z\omega}}}
\newcommand{\lii}{\lambda Z_\omega}

\def\doi{3 (3:6) 2007}
\lmcsheading%
{\doi}
{1--31}
{}
{}
{Oct.~13, 2006}
{Aug.~\phantom{0}6, 2007}
{}   

\begin{document}

\title{A Normalizing Intuitionistic Set Theory with Inaccessible Sets\rsuper*}

\author[W.~Moczyd\l owski]{Wojciech Moczyd\l owski}
\address{Department of Computer Science, Cornell University, Ithaca, NY, 14853, USA}
\email{wojtek@cs.cornell.edu}
\thanks{Partly supported by NSF grants DUE-0333526 and 0430161.}

\titlecomment{{\lsuper*}A version of this work is also available as a technical report
\cite{jatrinac2006}.}

\begin{abstract}
\noindent
We propose a set theory strong enough to interpret powerful type
theories underlying proof assistants such as LEGO and also possibly
Coq, which at the same time enables program extraction from its
constructive proofs. For this purpose, we axiomatize an impredicative
constructive version of Zermelo-Fraenkel set theory IZF with
Replacement and $\omega$-many inaccessibles, which we call \izfio. Our
axiomatization utilizes set terms, an inductive definition of
inaccessible sets and the mutually recursive nature of equality and
membership relations. It allows us to define a weakly-normalizing
typed lambda calculus corresponding to proofs in \izfio\ according to
the Curry-Howard isomorphism principle. We use realizability to prove
the normalization theorem, which provides a basis for program
extraction capability.
\end{abstract}

\maketitle

\section{Introduction}

Since the advent of proofs-as-programs paradigm, also called
propositions-as-types or Curry-Howard isomorphism, many systems with
program extraction capability have been built. Lego \cite{LP92}, Agda/Alfa
\cite{agda,alfa}, Coq
\cite{CoqManV8}, Nuprl \cite{nuprlbook}, Minlog \cite{BBS98} --- to name a few. Some are quite powerful ---
for example Coq can interpret an intuitionistic version of Zermelo's set
theory \cite{werner97}. With such power at hand, these systems have the 
potential of becoming very useful tools.

There is, however, one problem they all share, namely their foundational basis. In
order to use Coq or Nuprl, one has to master the ways of types,
a setting quite different from the set theory, the standard framework for doing
mathematics. A newcomer to this world, presented even with $\Pi$ and $\Sigma$
types emulating familiar universal and existential quantifiers, is likely to become
confused. The fact that the consistency of the systems is usually justified
by a normalization theorem in one form or other, does not make the matters easier. Even when set-theoretic
semantics is provided, it does not help much, given that the translation of
``the stamement $\forall x : \pl{nat}, \phi(x)$ is provable'' is ``the set
$\Pi_{n \in \nat} \SB{\phi[x:=n]}$ is inhabited'', instead of expected
``for all $x \in \nat$, $\phi(x)$ holds''. The systems which are not based on type theory 
share the problem of unfamiliar foundations. This is a serious shortcoming 
preventing the systems from becoming widely used, as the initial barrier to cross is set quite high.

In \cite{jacsl2006} we have made the first step to provide a solution to this problem, by
presenting a framework enabling extraction of programs from proofs, while using
the standard, natural language of set theory. That framework was based on the intuitionistic
set theory IZF with Replacement, called \izfr. Roughly speaking,
\izfr\ is what remains from Zermelo-Fraenkel set theory ZF after carefully
removing the excluded middle, while retaining the axioms of Power Set and unrestricted Separation.
The detailed exposition can be found in Section \ref{izfi}. For more
information on IZF and bibliography see \cite{scedrov85,beesonbook}.
We have defined a lambda calculus $\lambda Z$ corresponding to proofs in an
intensional version of \izfr\ and using realizability we have shown that $\lambda Z$ weakly normalizes.
By employing an inner model of extensional set theory, we have used the
normalization result to show that \izfr\ enjoys the standard properties of
constructive theories --- the disjunction, numerical existence, set existence and term existence properties 
(DP, NEP, SEP and TEP). These properties can be used to extract programs
from proofs \cite{chol}. All of them, apart from SEP, are essential to the
extraction process. However, even though \izfr\ is quite powerful, it is unclear if it is as
strong as type theories underlying the systems of Coq and LEGO, Calculus of Inductive Constructions (CIC) and Extended
Calculus of Constructions (ECC), as all known set-theoretical
interpretations use $\omega$-many strongly inaccessible
cardinals \cite{werner97,aczel98}.

We therefore axiomatize IZF with Replacement and $\omega$-many inaccessible
sets, which we call \izfio. Our axiomatization uses an inductive
definition of inaccessible sets. \izfio\ extended with excluded middle is equivalent to ZF with $\omega$-many
strong inaccessible cardinals. By utilizing the mutually recursive nature of
equality and membership relation, we avoid the need for the inner model and
define a lambda calculus $\lii$ corresponding directly to proofs in \izfio. 
We prove its normalization using realizability. As in \cite{jacsl2006}, normalization can be used to show DP, NEP, SEP and TEP. While DP and NEP have been proved for even stronger
theories in \cite{friedmanlarge}, our method is the first to provide the proof of TEP and SEP for intuitionistic set theory
with inaccessible sets. 

Inaccessible sets perform a similar function in a constructive setting to strongly inaccessible 
cardinals in the classical world and universes in type theories. They are ``large''
sets/types, closed under certain operations ensuring that they give rise to
models of set/type theories.  The closure conditions largely coincide in
both worlds and an inaccessible can be used to provide a set-theoretic intepretation of a
universe \cite{werner97,aczel98}. Both CIC and ECC have $\omega$-many
universes. By results of Aczel \cite{aczel98}, \izfio\ is strong enough to
interpret ECC. It is reasonable to expect that CIC could be interpreted too, as
the inductive types in CIC need to satisfy positivity conditions and
sufficiently strong inductive definitions are available in \izfio\ due to the presence of the
Power Set and unrestricted Separation axioms. Indeed, Werner's
set-theoretic interpretation \cite{werner97} of a large fragment of CIC uses only the
existence of inductively-defined sets in the set-theoretic universe to interpret
inductively-defined types. 

Our normalization result makes it possible to extract programs from proofs, 
using techniques described in \cite{chol}. Thus \izfio\ has all 
the proof-theoretic power of LEGO and likely Coq, uses familiar set-theoretic language and enables program
extraction from proofs. This makes it an attractive basis for a powerful and easy to use theorem prover.

This paper is mostly self-contained. We assume some familiarity with
set theory, proof theory and programming languages terminology, found for example 
in \cite{kunen,urzy,pierce}. 

The paper is organized as follows. In section \ref{ifol} we present
the intuitionistic first-order logic. We axiomatize IZF with Replacement and $\omega$-many
inaccessibles in sections \ref{izfi} and \ref{izfo}. In section \ref{lz} we
define the calculus $\lii$ and prove its standard properties. Realizability
is defined in section \ref{izfreal} and used to prove normalization
in section \ref{sectionnorm}. We describe related work in section
\ref{others}.

\section{Intuitionistic first-order logic}\label{ifol}

We start with a detailed presentation of the intuitionistic first-order logic
(IFOL). We use a natural deduction style of proof rules. The terms will be denoted by
letters $t, s, u$. The logical variables will be denoted by letters $a, b,
c, d, e, f$. The notation $\ov{a}$ denotes a finite sequence, treated as a set when
convenient. The $i$-th element of a sequence is denoted by $a_i$. We consider $\alpha$-equivalent
formulas equal. The capture-avoiding substitution is defined as usual; the
result of substituting $s$ for $a$ in a term $t$ is denoted by $t[a:=s]$. We
write $t[a_1, {\ldots} , a_n := s_1, {\ldots} , s_n]$ to denote the result
of substituting simultaneously $s_1, {\ldots} , s_n$ for $a_1, {\ldots} ,
a_n$. Contexts, denoted by $\Gamma$, are sets of formulas. 
The free variables of a formula $\phi$, denoted by $FV(\phi)$, are
defined as usual. The free variables of a context $\g$, denoted by $FV(\g)$, are
the free variables of all formulas in $\g$. The notation $\phi(\ov{a})$ means
that all free variables of $\phi$ are among $\ov{a}$. The proof rules are as follows:
\[
\infer{\g, \phi \p \phi}{} \qquad \infer{\gp \psi}{\gp \phi \to
\psi & \gp \phi} \qquad \infer{\gp \phi \to \psi}{\g, \phi \p \psi}
\]
\[
\infer{\gp \phi \land \psi}{\gp \phi & \gp \psi} \qquad \infer{\gp
\phi}{\gp \phi \land \psi} \qquad \infer{\gp \psi}{\gp \phi \land \psi}
\]
\[
\infer{\gp \phi \lor \psi}{\gp \phi} \quad \infer{\gp \phi \lor
\psi}{\gp \psi} \quad
\infer{\gp \vartheta}{\gp \phi \lor \psi & \g, \phi \proves \vartheta & \g, \psi
\proves \vartheta}
\]
\[
\infer[a \notin FV(\g)]{\gp \forall a.\ \phi}{\gp  \phi} \qquad
\infer{\gp \phi[a:=t]}{\gp \forall a.\ \phi} \qquad \infer{\gp \phi}{\gp \bot}
\]
\[
\infer{\gp \exists a.\ \phi}{\gp \phi[a:=t]} \qquad \infer[a \notin
FV(\g) \cup \{ \psi \}]{\gp \psi}{\gp \exists a.\ \phi &
\g, \phi \p \psi}
\]

Negation in IFOL is an abbreviation: $\lnot \phi \equiv \phi \to
\bot$. So is the symbol $\iffl$: $\phi \iffl \psi \equiv (\phi \to \psi \land
\psi \to \phi)$. Note that IFOL does not contain equality. The excluded middle rule added to IFOL makes it equivalent 
to the classical first-order logic without equality.

\begin{lemma}\label{formsubst}
For any formula $\phi$, $\phi[a:=t][b:=u[a:=t]] = \phi[b:=u][a:=t]$, for $b \notin FV(t)$. 
\end{lemma}
\begin{proof}
Straightforward structural induction on $\phi$. 
\end{proof}

\section{\izfi}\label{izfi}

In this section we introduce our first approximation to \izfr, called
\izfi, which is \izfr\ from \cite{jacsl2006} extended with the axioms postulating the existence of inaccessible sets. 
We start by presenting the axioms of \izfr. It is a first-order theory. When extended with excluded middle, it is equivalent to ZF.
The signature consists of two binary relational symbols $\in, =$ and function symbols used in the axioms below. The symbols 
$0$ and $S(a)$ are abbreviations for $\emptyset$ and $\bigcup \{ a, \{ a, a \} \}$. Bounded quantifiers and the quantifier $\exists !a$ (there exists exactly one $a$) are
also abbreviations defined in the standard way.

\begin{enumerate}[$\bullet$]
\item (EXT) $\forall a, b.\ a = b \iffl \forall c.\ c \in a \iffl c \in b$
\item (L${}_\phi$) $\forall a, b, \ov{f}.\ a = b \land \phi(a, \ov{f}) \to \phi(b,
\ov{f})$
\item (EMPTY) $\forall c.\ c \in \emptyset \iffl \bot$
\item (PAIR) $\forall a, b \forall c.\ c \in \{ a, b \} \iffl c = a \lor c = b$
\item (INF) $\forall c.\ c \in \omega \iffl c = 0 \lor \exists b \in \omega.\ c =
S(b)$
\item (SEP${}_{\sphi}$)
$\forall \ov{f} \forall a \forall
c.\ c \in S_{\sphi}(a, \ov{f}) \iffl c \in a \land
\phi(c, \ov{f})$
\item (UNION): $\forall a \forall c.\ c \in \bigcup a \iffl \exists b \in
a.\ c \in b$
\item (POWER) $\forall a \forall c.\ c \in P(a) \iffl \forall b.\ b \in c \to b \in a$
\item (REPL${}_{\rphi} $) $\forall \ov{f}, a
\forall c.\ c \in R_{\rphi}(a, \ov{f}) \iffl
(\forall x \in a \exists! y.\ \phi(x, y, \ov{f})) \land (\exists x \in a.\ \phi(x, c, \ov{f}))$
\item (IND${}_{\iphi}$) $\forall \ov{f}. (\forall a. (\forall b \in
a.\ \phi(b, \ov{f})) \to \phi(a, \ov{f})) \to \forall a.\ \phi(a, \ov{f})$
\end{enumerate}

The axioms (SEP${}_\phi$), (REPL${}_{\phi}$), (IND${}_\phi$) and (L${}_\phi$) are axiom schemas ---
there is one axiom for each formula $\phi$. Note that there are terms $S_\phi$ and $R_\phi$ for 
each instance of the Separation and Replacement axioms. Formally, terms and formulas are defined by mutual induction:
\[
\phi ::= t \in t\ |\ t = t\ | {\ldots} \qquad
t ::= a\ |\ \{ t, t \}\ |\ \ S_{\sphi}(t, \ov{t})\ |\ R_{\rphi}(t, \ov{t})\ | {\ldots}
\]
The axioms (EMPTY), (PAIR), (INF), (SEP${}_{\phi}$), (UNION), (POWER) and (REPL$_{\phi}$)
all assert the existence of certain classes and have the same form: $\forall 
\ov{a}. \forall c.\ c \in t_A(\ov{a}) \iffl \phi_A(c, \ov{a})$, where $t_A$ is a 
function symbol and $\phi_A$ a corresponding formula
for the axiom (A). For example, for (POWER), $t_{\mathit{POWER}}$ is $P$ and
$\phi_{\mathit{POWER}}$ is $\forall b.\ b \in c
\to b \in a$. We reserve the notation $t_A$ and $\phi_A$ to denote the term and
the corresponding formula for the axiom (A).

The terms $S_{\phi}(t, \ov{t})$ and $R_{\rphi}(t, \ov{t})$ could
be displayed as $\{ c \in t\ |\ \phi(c, \ov{t}) \}$ and
$\{ c\ |\ (\forall x \in t \exists! y \phi(x, y, \ov{t})) \land (\exists x
\in t.\
\phi(x, c, \ov{t})) \}$, respectively.

\subsection{On the axioms of \izfr}

\subsubsection{The Leibniz axiom}

The Leibniz axiom (L${}_\phi$) is usually not present among the axioms of set
theories, as it is assumed that logic contains equality and the axiom is 
a proof rule. We include (L${}_\phi$) among the axioms of \izfr, because
there is no obvious way to add it to intuitionistic logic in the Curry-Howard isomorphism context,
as its computational content is unclear.

\subsubsection{The Replacement axiom}

A more familiar formulation of Replacement could be: ``For all $\ov{F}, A$, if for all $x \in A$ there is exactly one $y$
such that $\phi(x, y, \ov{F})$ holds, then there is a set $D$ such that
$\forall x \in A \exists y \in D.\ \phi(x, y, \ov{F})$ and for all $d \in D$ there is $x
\in A$ such that $\phi(x, d, \ov{F})$''. 
Let this formulation of Replacement be called (REPL0$_{\rphi}$), let
($R_\phi$) be the term-free statement of our Replacement axiom, that is:
\[
(R_\phi) \equiv \forall \ov{f}, a \exists !d.\ \forall c.\ c
\in d \iffl (\forall x \in a \exists! y.\ \phi(x, y, \ov{f})) \land (\exists x \in a.\ \phi(x, c, \ov{f}))
\]
and let IZ denote \izfr\ without the Replacement axiom and corresponding function symbols.
To justify our definition of Replacement, we prove the following two lemmas:
\begin{lemma}\label{repl0}
IZ $\p$ (R$_{\rphi}$) $\to $(REPL$0_{\rphi}$).
\end{lemma}
\begin{proof}
Assume (R$_{\rphi}$), take any $\ov{F}, A$ and suppose that for all $x \in A$ there is exactly one $y$ such that $\phi(x,
y, \ov{F})$. Let $D$ be the set we get by applying $(R_\phi)$. Take any $x \in
A$, then there is $y$ such that $\phi(x, y, \ov{F})$, so $y \in D$.
Moreover, if $d \in D$ then there is $x \in A$
such that $\phi(x, d, \ov{F})$. This shows (REPL0$_{\phi}$). 
\end{proof}
\begin{lemma}\label{repl}
IZ $\p$ (REPL0$_{\phi}$) $\to $(R$_{\rphi}$).
\end{lemma}
\begin{proof}
Assume (REPL0$_{\rphi}$), take any $\ov{F}, A$ and consider the set 
\[
B \equiv \{
a \in A\ |\ \forall x \in A \exists !y.\ \phi(x, y, \ov{F}) \}.
\]
Then for all $b \in B$ there is exactly one $y$ such that $\phi(b, y, \ov{F})$. Use
(REPL0$_{\rphi}$) to get a set $D$. Then $D$ is the set we are looking for. Indeed, 
if $d \in D$, then there is $b \in B$ such that $\phi(b, d, \ov{F})$ and so
by the definition of $B$, $\forall x \in A \exists !y. \ \phi(x, y, \ov{F})$
and $b \in A$. On the other hand, take any $d$ and suppose that $\forall x \in
A \exists !y.\ \phi(x, y, \ov{F})$ and there is $x \in A$ such that $\phi(x,
d, \ov{F})$. Then $x \in B$, so there is $y' \in D$ such that $\phi(x,
y', \ov{F})$. But $y'$ must be equal to $d$, so $d \in D$. As it is trivial
to see that $D$ is unique, the claim follows.
\end{proof}

\ignore{
This argument justifies the definitional extension of \izfro\ with function symbols $R_{\phi}(a,
\ov{f})$, where $\phi$ is in the language of \izfro. For any formula $\psi$ using
these new terms, there is an equivalent formula $\psi'$ in the language of
\izfro\ and their equivalence can be shown in \izfro.

Now by a simple inductive argument we can show that for any natural number
$n$ and formula $\phi$ of replacement depth $n$, \izfro\ can be definitionally extended by the replacement terms $R_{\phi}$,
The base case, where $n = 0$, has already been shown. For the inductive
step, let $\phi$ contain replacement terms $r_1,
{\ldots} , r_k$ of depth $n$. We need to show that the class $A = \{ z\ |\ \forall x \in A \exists !y.\
\phi(x, y, \ov{f}, r_1, {\ldots} , r_k) \land \exists x \in A.\ \phi(x, y,
\ov{f}, r_1, {\ldots} , r_k)\}$ is a set. By the inductive hypothesis, there is a
formula $\phi'(x, y, \ov{f})$ such that \izfro $\p \phi \iffl \phi'$. 
Therefore $A = \{ z\ |\ \forall x \in A \exists !y.\
\phi'(x, y, \ov{f}) \land \exists x \in A.\ \phi'(x, y, \ov{f})\}$. 
As $\phi'$ does not contain any replacement terms, the argument we used for the base case shows the claim.

By combining all these definitional extensions together, we get exactly \izfr.

}

\subsubsection{The terms of \izfr}

The original presentation of IZF with Replacement presented in \cite{myhill73} is
term-free. Let us call it \izfro. We will now show that \izfr\ is a
definitional extension of \izfro. 

In \izfro\ for each axiom (A) among the Empty Set, Pairing, Infinity, Separation,
Replacement, Union and Power Set axioms, we can derive $\forall \ov{a} \exists !d \forall c.\ c \in d \iffl \phi_A(c, \ov{a})$, 
using Lemma \ref{repl} in case of the Replacement axiom. We therefore
definitionally extend \izfro, by introducing for each such (A) the corresponding new function symbol $t_A(\ov{a})$ along with the
defining axiom $\forall \ov{a} \forall c.\ c \in t_A(\ov{a}) \iffl \phi_A(c,
\ov{a})$.

We then need to provide the Separation and Replacement function symbols $R_{\phi}$
and $S_\phi$, where $\phi$ may contain the new terms. To fix our attention, consider the Separation axiom.
For some function symbol $S_\phi$, we need to have:
\[
\forall \ov{f}, a \forall c.\ c \in S_\phi(a, \ov{f}) \iffl c \in a \land
\phi(c, \ov{f})
\]
As all terms present in $\phi$ were introduced via a definitional extension
of \izfro, there is a term-free formula $\phi'$ equivalent to $\phi$. We
therefore have:
\[
\forall \ov{f}, a \forall c.\ c \in S_{\phi'}(a, \ov{f}) \iffl c \in a \land
\phi'(c, \ov{f})
\]
and consequently:
\[
\forall \ov{f}, a \forall c.\ c \in S_{\phi'}(a, \ov{f}) \iffl c \in a \land
\phi(c, \ov{f})
\]
We define $S_{\phi}$ to be $S_{\phi'}$. Similarly, we can define $R_{\phi}$
to be $R_{\phi'}$. After iterating this process $\omega$-many times, we obtain all instances of
terms and axioms (A) present in \izfr. 

It remains to derive the Leibniz and $\in$-Induction axioms for formulas with
terms. For the Leibniz axiom, take any $A, B, \ov{F}$ and suppose $A = B$ and
$\phi(A, \ov{F})$. Then there is a term-free formula $\phi'$ equivalent to
$\phi$, so also $\phi'(A, \ov{F})$. By the Leibniz axiom in \izfro,
$\phi'(B, \ov{F})$, so also $\phi(B, \ov{F})$. 

For the $\in$-Induction axiom, take any $\ov{F}$ and suppose:
\[
\forall a.\ (\forall b \in a.\ \phi(b, \ov{F})) \to \phi(a, \ov{F})
\]
Taking $\phi'$ to be the term-free formula equivalent to $\phi$, we get:
\[
\forall a.\ (\forall b \in a.\ \phi'(b, \ov{F})) \to \phi'(a, \ov{F})
\]
By $\in$-Induction in \izfro, we get $\forall a.\ \phi'(a, \ov{F})$, thus
also $\forall a.\ \phi(a, \ov{F})$.

\subsection{Inaccessible sets}

To extend \izfr\  with inaccessible sets, we add a family of axioms
(INAC${}_i$) for $i \gt 0$. We call the resulting theory \izfi. 
The axiom (INAC${}_i$) asserts the existence of
the $i$-th inaccessible set, denoted by a new constant symbol $V_i$, and is defined as follows:

\[
(\mbox{INAC}_i)\ \forall c.\ c \in V_i \iffl \phiinaca(c, V_i) \land
\forall d.\ \phiinacb(d) \to c \in d
\]
Following the conventions set up for \izfr,
$\phi_{INAC_i}(c)$ is $\phiinaca(c, V_i) \land \forall d.\ \phiinacb(d) \to c
\in d$. The formula $\phiinaca(c, d)$ intuitively sets up conditions for $c$
being a member of $V_i$, while $\phiinacb(d)$ says what it means for $d$
to be inaccessible. To streamline the definition, we set $V_0$ to abbreviate $\omega$.
\begin{definition}\label{dinac}
The formula $\phiinaca(c, V_i)$ for $i \gt 0$ is a disjunction of the
following five clauses:
\begin{enumerate}[(1)]
\item \label{i0} $c = V_{i-1}$
\item there is $a \in V_i$ such that $c \in a$.
\item there is $a \in V_i$ such that $c$ is a union of $a$.
\item there is $a \in V_i$ such that $c$ is a power set of $a$.
\item there is $a \in V_i$ such that $c$ is a function from $a$ to
$V_i$. 
\end{enumerate}
\end{definition}
\begin{definition}\label{d2}
The formula $\phiinacb(d)$ for $i \gt 0$ is a conjunction of the
following five clauses:
\begin{enumerate}[(1)]
\item $V_{i-1} \in d$.
\item $\forall e, f.\ e \in d \land f \in e \to f \in d$.
\item $\forall e \in d.\ \bigcup e \in d$.
\item $\forall e \in d.\ P(e) \in d$.
\item $\forall e \in d.\ \forall f \in e \to d.\ f \in d$, where $e \to d$
denotes the set of all functions from $e$ to $d$. 
\end{enumerate}
\end{definition}

Briefly, the $i$-th inaccessible set is the smallest transitive set containing $V_{i-1}$ and
closed under unions, power sets and taking functions from its elements into
itself. It is easy to see that \izfi + EM is equivalent to ZF with 
$\omega$-many strongly inaccessible cardinals. For a theory $T$, let $M(T)$
denote a sentence ``$T$ has a model''. To show that the set $V_i$ defined by (INAC${}_i$) behaves as an
inaccessible set in \izfi\ we prove:

\begin{thm}[\izfi]
For all $i \gt 0$, $V_i \models $\izfr + M(\izfr) + M(\izfr + M(\izfr))  + {\ldots} ($i$ times). 
\end{thm}
\begin{proof}
\ifthenelse{\boolean{long}}
{
To make this claim precise, we first need to provide the interpretation of terms and
relational symbols in $V_i$. Equality and set membership are induced from
$V$. For the function symbols, let $A, \ov{F} \in V_i$. 
\begin{itemize}
\item $\omega^{V_i}$ is $\omega$, available in $V_i$
\item $P(A)^{V_i}$ is $P(A)$, available in $V_i$
\item $(\bigcup\ A)^{V_i}$ is $\bigcup A$.
\item $\{ x \in A\ |\ \phi(x, \ov{F}) \}^{V_i}$ is\footnote{This is slightly
informal, as at the moment the satisfiability relation has not been defined
yet. Fully formal treatment would define satisfiability and interpretation of 
terms by mutual induction on the definition of terms and formulas.} 
$\{ x \in A\ |\ V_i \models
\phi(x, \ov{F}) \}$. This is a member of $P(A)$, so transitivity of $V_i$ ensures it's a member of $V_i$. 
\item $\{ y\ |\ \forall x \in A \exists !y.\ \phi(x, y, \ov{F}) \land
\exists x \in A.\ \phi(x, y, \ov{F}) \}^{V_i}$ is $\{ y\ |\
\forall x \in A \exists !y \in V_i.\ V_i \models \phi(x, y, \ov{F}) \land
\exists x \in A.\ V \models \phi(x, y, \ov{F})$.
We need to show that it's in $V_i$. Take $B = \{ x \in A\ |\ \forall x \in A
\exists !y \in V_i.\ V_i \models \phi(x, y, \ov{F}) \}$. Then $B \in V_i$. 
Now take $C = \{ (x, y) \in B \times V_i\ |\ V_i \models \phi(x, y, \ov{F}) \}$. Then for all
$x \in B$ there is exactly one $y$ such that $(x, y) \in C$. Thus $C \in B
\to V_i$. Therefore $C \in V_i$. So is $ran(C)$. Suppose $y \in ran(C)$. Then 
there is $x \in B$ such that $V_i \models \phi(x, y, \ov{F})$, so also
$\forall x \in A \exists !y \in V_i.\ V_1 \models \phi(x, y, \ov{F})$, so $y
\in \{ y\ |\
\forall x \in A \exists !y \in V_i.\ V_i \models \phi(x, y, \ov{F}) \land
\exists x \in A.\ V \models \phi(x, y, \ov{F})$. On the other hand, suppose
$y \in \{ y\ |\
\forall x \in A \exists !y \in V_i.\ V_i \models \phi(x, y, \ov{F}) \land
\exists x \in A.\ V \models \phi(x, y, \ov{F})$. Then there is $x \in A$
such that $V \models \phi(x, y, \ov{F})$ and $\forall x \in A \exists !y \in
V_1.\ V_1 \models \phi(x, y, \ov{F})$, so $x \in B$ and $(x, y) \in C$, so
$y \in ran(F)$. Thus this class is indeed in $V_i$. 
\end{itemize}
We proceed axiom by axiom:
\begin{itemize}
\item (EXT), (L) are immediate, as $\in$ and $=$ are absolute. Note also 
that $V_i$ is transitive. 
\item (EMPTY) Note that by clause \ref{i0} of the definition \ref{dinac},
$\omega$ is in $V_1$. Since $\emptyset$ is a subset of $\omega$, by clause 
\item (INF) We know that $\omega$ is in $V_i$. Since the defining formula is
$\Delta_0$ and hence absolute, we get the claim.
\item SEP${}_{\phi(a, \ov{f})}$. Take $A, \ov{F} \in V_i$ and let $B = \{ c
\in a\ |\ V_i \models \phi(c, \ov{f}) \}$. Since $B \subseteq A$, $B \in
V_i$. If $c \in B$, then $C \in A$ and $\phi^{V_i}(C, \ov{F})$. And the
other way round, too.
\item (UNION), (POWER) Straightforward.
\item (IND${}_{\phi(a, \ov{f})}$) We want to show:
\[
\forall \ov{f} \in V_i.\ (\forall a \in V_i.\ 
(\forall b \in V_i.\ b \in a \to \phi^{V_i}(b, \ov{f})) \to \phi^{V_i}(a, \ov{f})) \to
\forall a \in V_i. \phi^{V_i}(a, \ov{f})
\]
Take any $\ov{F} \in V_i$. It suffices to show that:
\[
(\forall a.\ a \in V_i \to 
(\forall b.\ b \in V_i \to b \in a \to \phi^{V_i}(b, \ov{F})) \to
\phi^{V_i}(a, \ov{F})) \to
\forall a.\ a \in V_i \to \phi^{V_i}(a, \ov{F})
\]
This is equivalent to:
\[
(\forall a.\ (\forall b.\ b \in a \to b \in V_i \to \phi^{V_i}(b, \ov{F})) \to
a \in V_i \to \phi^{V_i}(a, \ov{F})) \to
\forall a.\ a \in V_i \to \phi^{V_i}(a, \ov{F})
\]
But this the instance of the induction axiom for the formula $a \in V_i \to
\phi^{V_i}(a, \ov{f})$.
\item (REPL) We will show the standard version of replacement and use Lemma
to get the claim. Take any $\ov{F}, A \in V_i$ and suppose that $\forall x
\in A \exists !y \in V_i.\ \phi^{V_i}(x, y, \ov{F})$. Consider $\{ (x, y)
\in A \times V_i\ |\ \phi^{V_i}(x, y, \ov{F}) \}$. This is a function from
$a \to V$, so its range, which is $\{ y \in V_i\ |\ \exists x \in A.\ 
\phi^{V_i}(x, y, \ov{F}) \}$ is in $V_i$. It's trivial to check that it has
the required properties. 
\end{itemize}
For $i \gt 0$, we have to show that $V_{i+1} \models IZF + Con(IZF) +
{\ldots} + Con^i(IZF)$. The proof that $V_{i+1} \models IZF$ is the same. We
know that $V_i \in V_{i+1}$. By IH, $V_i \models IZF + Con(IZF) + {\ldots} + 
Con^{i-1}(IZF)$, thus $V_{i+1} \models Con(IZF+Con(IZF) + {\ldots}  +
Con^{i-1}(IZF))$.}
{
By Clause 2 in the Definition \ref{dinac}, $V_1$ is transitive, so the equality
and membership relations are absolute. Clause 1 gives us $\omega \in V_1$ and since its
definition is $\Delta_0$,  $V_1 \models $(INF). Clauses 3 and 4 provide the
(UNION) and (POWER) axioms. Transitivity then gives (SEP) and (PAIR), while
Clause 5, thanks to Lemma \ref{repl}, gives (REPL$_{\phi}$). The existence
of the empty set follows by (INF) and (SEP). For the Induction axiom, we
need to show:
\[
\forall \ov{f} \in V_i.\ (\forall a \in V_i.\ 
(\forall b \in V_i.\ b \in a \to \phi^{V_i}(b, \ov{f})) \to \phi^{V_i}(a, \ov{f})) \to
\forall a \in V_i. \phi^{V_i}(a, \ov{f})
\]
Take any $\ov{F} \in V_i$. It suffices to show that:
\[
(\forall a.\ a \in V_i \to 
(\forall b.\ b \in V_i \to b \in a \to \phi^{V_i}(b, \ov{F})) \to
\phi^{V_i}(a, \ov{F})) \to
\forall a.\ a \in V_i \to \phi^{V_i}(a, \ov{F})
\]
This is equivalent to:
\[
(\forall a.\ (\forall b.\ b \in a \to b \in V_i \to \phi^{V_i}(b, \ov{F})) \to
a \in V_i \to \phi^{V_i}(a, \ov{F})) \to
\forall a.\ a \in V_i \to \phi^{V_i}(a, \ov{F})
\]
But this is the instance of the induction axiom for the formula $a \in V_i \to
\phi^{V_i}(a, \ov{f})$.

Thus $V_1 \models $\izfr. Since $V_1 \in V_2$, 
$V_2 \models $ \izfr + M(\izfr). Since $V_2 \in V_3$, $V_3
\models $\izfr + M(\izfr + M(\izfr)). Proceeding in this manner by induction we get the
claim.
}
\end{proof}

\section{\izfio}\label{izfo}

We now present our final axiomatization of IZF with Replacement and
inaccessible sets, which we call \izfio. The advantage of this axiomatization
over the previous one is that equality and membership are defined in terms of
each other, instead of being taken for granted and axiomatized with
Extensionality and Leibniz axioms. This trick, which amounts to
interpreting an extensional set theory in an intensional one, has  already
been used by Friedman in \cite{friedmancons}. As we shall see later, this makes it
possible to prove a normalization theorem directly for the theory, thus avoiding the
need for the detour via the class of transitively-L-stable sets used in
\cite{jacsl2006}. 

The signature of \izfio\ consists of three relational symbols: $\ini, \ino, \eqo$ and
terms of \izfi. The axioms of \izfio\ are as follows:

\begin{enumerate}[$\bullet$]
\item (IN) $\forall a, b.\ a \ino b \iffl \exists c.\ c \ini b \land a \eqo
c$
\item (EQ) $\forall a, b.\ a \eqo b \iffl \forall d.\ (d \ini a \to d \ino b)
\land (d \ini b \to d \ino a)$
\item (IND${}_{\phi}$) $\forall \ov{f}. (\forall a. (\forall b \ini
a. \phi(b, \ov{f})) \to \phi(a, \ov{f})) \to \forall a. \phi(a, \ov{f})$
\item (A) $\forall \ov{a}.\ \forall c.\ c \ini t_A(\ov{a}) \iffl \phi_A(c,
\ov{a})$, for (A) being one of (EMPTY), (PAIR), (INF), (SEP${}_\phi$), (UNION),
(POWER), (REPL${}_\phi$), (INAC${}_i$). For example, the Power Set axiom has a
form: $\forall a \forall c.\ c \ini P(a) \iffl \forall b.\ b \in c \to b \in a$. 
\end{enumerate}

The extra relational symbol $\ini$ intuitively denotes the intensional membership relation. Note that neither the Leibniz axiom (L$_\phi$) nor
the extensionality axiom are present.
We will show, however, that they can be derived and that this 
axiomatization is as good as \izfi. From now on in this section, we
work in \izfio. The following sequence of lemmas establishes that equality
and membership behave in the correct way. Statements similar in spirit are also proved 
in the context of Boolean-valued models. Our treatment slightly simplifies the standard 
presentation by avoiding the need for mutual induction. 

\begin{lemma}\label{eqrefl}
For all $a$, $a \eqo a$.
\end{lemma}
\begin{proof}
By $\in$-induction on $a$. Take any $b \ini a$. By the inductive hypothesis, $b = b$, so also $b \in a$. 
\end{proof}

\begin{corollary}\label{cor1}
If $a \ini b$, then $a \ino b$. 
\end{corollary}

\begin{lemma}\label{eqsymm}
For all $a, b$, if $a \eqo b$, then $b \eqo a$.
\end{lemma}
\begin{proof}
Straighforward.
\end{proof}
\begin{lemma}\label{eqtrans}
For all $b, a, c$, if $a = b$ and $b = c$, then $a = c$. 
\end{lemma}
\begin{proof}
By $\in$-induction on $b$. First take any $d \ini a$. By $a = b$, $d \in b$,
so there is $e \ini b$ such that $d = e$. By $b = c$, $e \in c$, so there is
$f \ini c$ such that $e = f$. By the inductive hypothesis for $e$, $d = f$,
so $d \in c$.

The other direction is symmetric and proceeds from $c$ to $a$. Take any $d
\ini c$. By $b = c$, $d \in b$, so there is $e \ini b$ such that $d = e$. By
$a = b$, $e \in a$, so there is $f \ini a$ such that $e = f$. The inductive
hypothesis gives the claim.
\end{proof}

\begin{lemma}\label{lei0}
For all $a, b, c$, if $a \in c$ and $a = b$, then $b \in c$. 
\end{lemma}
\begin{proof}
Since $a \in c$, there is $d \ini c$ such that $a = d$. By previous lemmas
we also have $b = d$, so $b \in c$. 
\end{proof}

\begin{lemma}\label{ext0}
For all $a, b, d$, if $a = b$ and $d \in a$, then $d \in b$.
\end{lemma}
\begin{proof}
Suppose $d \in a$, then there is $e$ such that $e \ini a$ and $d = e$. By
$a = b$, $e \in b$. By Lemma \ref{lei0}, $d \in b$. 
\end{proof}

\begin{lemma}[Extensionality]\label{ext}
If for all $d$, $d \in a$ iff $d \in b$, then $a = b$. 
\end{lemma}
\begin{proof}
Take any $d \ini a$. By Corollary \ref{cor1} $d \in a$, so by Lemma \ref{ext0} also $d \in b$.
The other direction is symmetric. 
\end{proof}

We would like to mention that all the lemmas above have been verified by the
computer, by a toy prover we wrote to experiment with \izfio.

\begin{lemma}[The Leibniz axiom]\label{lei}
For any term $t(a, \ov{f})$ and formula $\phi(a, \ov{f})$ not containing $\ini$, if $a \eqo b$, then $t(a, \ov{f}) \eqo t(b, \ov{f})$
and $\phi(a, \ov{f}) \iffl \phi(b, \ov{f})$. 
\end{lemma}
\proof 
Straightforward mutual induction on generation of $t$ and $\phi$. 
We show some representative cases.
Case $t$ or $\phi$ of:
\begin{enumerate}[$\bullet$]
\ignore
{
\item $a, f_i, \omega, V_i, \emptyset$. The claim is immediate.
\item $\{ t_1(a, \ov{f}), t_2(a, \ov{f}) \}$. By the inductive hypothesis, $t_1(a, \ov{f}) =
t_1(b, \ov{f})$ and $t_2(a, \ov{f}) = t_2(b, \ov{f})$. Suppose $c \ini \{ t_1(a,
\ov{f}), t_2(a, \ov{f}) \}$. To fix our attention, let $c \eqo t_1(a,
\ov{f})$. Then $c \eqo t_1(b, \ov{f})$ and thus $c \ini \{ t_1(b, \ov{f}),
t_2(b, \ov{f}) \}$, so by Corollary \ref{cor1} we get the claim. The other
direction is symmetric. 
}
\item $\bigcup t_1(a)$. If $c \ini \bigcup t_1(a)$, then for some $d$, $c \ino d \ino t_1(a)$.
By the inductive hypothesis $t_!(a) \eqo t_1(b)$, so by Lemma \ref{ext0} $d \ino t_1(b)$, so $c \ini
\bigcup t_1(b)$ and by Corollary \ref{cor1} also $c \ino \bigcup t_1(b)$.
The other direction is symmetric and by the (EQ) axiom we get $t(a) = t(b)$.
\item $S_\phi(t_1(a), \ov{u}(a))$. If $c \ini S_\phi(t_1(a), \ov{u}(a))$, then $c
\ino t_1(a)$ and $\phi(c,
\ov{u}(a))$. By the inductive hypothesis, $t_1(a) = t_1(b)$, 
$\ov{u}(a) = \ov{u}(b)$, and thus $\phi(c, \ov{u}(b))$ and $c \in t_1(b)$, so
$c \ini S_{\phi}( t_1(b), \ov{u}(b))$ and also $c \in S_{\phi}( t_1(b),
\ov{u}(b))$. 
\item $t(a) \ino s(a)$. By the inductive hypothesis, $t(a) \eqo t(b)$ and $s(a) \eqo s(b)$. 
Thus by Lemma \ref{ext0} $t(a) \ino s(b)$ and by Lemma \ref{lei0} $t(b) \ino s(b)$.
\item $\forall c.\ \phi(c, a, \ov{f})$. Take any $c$, we have $\phi(c, a,
\ov{f})$, so by inductive hypothesis $\phi(c, b, \ov{f})$, so $\forall c.\
\phi(c, b, \ov{f})$.\qed
\end{enumerate}

\begin{lemma}\label{rest}
For any term $t_A(\ov{a})$, $c \ino t_A(\ov{a})$ iff $\phi_A(c, \ov{a})$.
\end{lemma}
\begin{proof}
The right-to-left direction follows immediately by Corollary \ref{cor1}. For the
left-to-right direction, suppose $c \in t_A(\ov{a})$. Then there is $d$ such
that $d \ini t_A(\ov{a})$ and $c = d$. Therefore $\phi_A(d, \ov{a})$ holds
and by the Leibniz axiom we also get $\phi_A(c, \ov{a})$. 
\end{proof}

\begin{lemma}
For any axiom $A$ of \izfi, \izfio $\p A$.
\end{lemma}
\begin{proof}
Lemmas \ref{ext}, \ref{lei} and \ref{rest} show the claim for all the axioms
apart from (IND$_\phi$). So suppose $\forall a.\ (\forall b \ino a.\ \phi(b, \ov{f})) \to
\phi(a, \ov{f})$. We need to show $\forall a.\ \phi(a, \ov{f})$. We proceed by
$\ini$-induction on $a$. It suffices to show $\forall c.\ (\forall d \ini c.\
\phi(d, \ov{f})) \to \phi(c, \ov{f})$. Take any $c$ and suppose $\forall d \ini
c.\ \phi(d, \ov{f})$. We need to show $\phi(c, \ov{f})$. Take $a$ to be $c$ in
the assumption, so it suffices to show that $\forall b \ino c.\ \phi(b,
\ov{f})$. Take any $b \ino c$. Then there is $e \ini c$ such that $e \eqo b$. 
By the inductive hypothesis $\phi(e, \ov{f})$ holds and hence by the Leibniz
axiom we get $\phi(b, \ov{f})$, which shows the claim. 
\end{proof}

\begin{corollary}\label{izf01}
If \izfi $\p \phi$, then \izfio $\p \phi$.
\end{corollary}

\begin{lemma}\label{izf02}
If \izfio $\p \phi$ and $\phi$ does not contain $\ini$, then \izfi $\p \phi$.
\end{lemma}
\begin{proof}
Working in \izfi\ simply interpret $\ini$ as $\in$ to see that all axioms
of \izfio\ are valid and that if \izfio $\p \phi$, then \izfi $\p
\phi[\ini := \in]$. 
\end{proof}

Therefore \izfio\ is a legitimate axiomatization of IZF with Replacement and
inaccessible sets. From now on the names of the axioms refer to the
axiomatization of \izfio. 

\section{The $\lii$ calculus}\label{lz}

We now introduce a lambda calculus $\lii$ for \izfio, based on the Curry-Howard isomorphism
principle. The part of $\lii$ corresponding to the first-order logic is
essentially $\lambda P_1$ from \cite{urzy}. The rest of the calculus, apart
from clauses corresponding to (IN), (EQ) and (INAC$_{i}$) axioms, is
identical to $\lambda Z$ from \cite{jacsl2006}.
\subsection{The terms of $\lii$}

The lambda terms in $\lii$ will be denoted by letters $M, N, O, P$. There
are two kinds of lambda abstraction in $\lii$, one corresponding to the
proofs of implication, the other to the proofs of universal quantification.
We use separate sets of variables for these abstractions and call them
propositional and first-order variables, respectively. Letters $x, y, z$
will be used for the propositional variables and letters $a, b, c$ for the first-order
variables. Letters $t, s, u$ are reserved for \izfio\  terms. The types in the system
are \izfio\ formulas. The terms are generated by the following abstract
grammar:
\[
M ::= x\ |\ M\ N\ |\ \lambda a.\ M\ |\ \lambda x : \phi.\ M\ |\ \INL(M)\ |\
\INR(M)\ |\ \FST(M)\ | \ \SND(M)
\]
\[
[t, M]\ |\ M\ t\ |\ <M, N>\ |\ \CASE(M, x : \phi.\ N, x : \psi.\ O)\ |\
\MAGIC(M)\ |\ \LET\ [a, x : \phi] := M\ \IN\ N
\]
\[
\IND_{\phi(a, \ov{b})}(M, \ov{t})\ |\ \pl{inac}_i\pl{Prop}(t, M)\ |\ \pl{inac}_i\pl{Rep}(t, M)
\]
\[
\pl{inProp}(t, u, M)\ |\ \pl{inRep}(t, u, M)\ |\ \pl{eqProp}(t, u, M)\ |\ \pl{eqRep}(t, u, M)
\]
\[
\pl{pairProp}(t, u_1, u_2, M)\ |\ \pl{pairRep}(t, u_1, u_2, M)\ |\ \pl{unionProp}(t, u, M)\ | \ \pl{unionRep}(t, u, M)
\]
\[
\pl{sep}_{\phi(a, \ov{f})}\pl{Prop}(t, u, \ov{u}, M)\ |\ \pl{sep}_{\phi(a,
\ov{f})}\pl{Rep}(t, u, \ov{u}, M)\ |\ \pl{powerProp}(t, u, M)\ | \ \pl{powerRep}(t, u, M)
\]
\[
\pl{infProp}(t, M)\ | \ \pl{infRep}(t, M)\ |\ \pl{repl}_{\phi(a, b,
\ov{f})}\pl{Prop}(t, u, \ov{u}, M)\ |\ \pl{repl}_{\phi(a, b, \ov{f})}\pl{Rep}(t, u, \ov{u}, M)
\]

The \pl{ind} terms correspond to the (IND) axiom, \pl{Prop} and
\pl{Rep} terms correspond to the respective axioms of \izfi and the rest of
the terms corresponds to the rules of IFOL. The exact nature of the
correspondence will become clear in Section \ref{types}. 
To avoid listing all of them repeatedly, we adopt a convention of using \pl{axRep} and \pl{axProp} terms to tacitly
mean all \pl{Rep} and \pl{Prop} terms, for \pl{ax} being one of \pl{in},
\pl{eq}, \pl{pair}, \pl{union}, \pl{sep}, \pl{power}, \pl{inf}, \pl{repl} and 
\pl{inac_i}, unless we list some of them separately. 
With this convention in mind, we can summarize the definition of the \pl{Prop} and \pl{Rep} terms as:
\[
\pl{axProp}(t, \ov{u}, M)\ |\ \pl{axRep}(t, \ov{u}, M), 
\]
where the number of terms in the sequence $\ov{u}$ depends on the particular
axiom. 

The free variables of a lambda term are defined as usual, taking into
account that variables in $\lambda$, \pl{case} and \pl{let} terms bind respective
terms. The relation of $\alpha$-equivalence is defined taking this information into account. We consider $\alpha$-equivalent terms equal.
We denote all free variables of a term $M$ by $FV(M)$ and the free first-order
variables of a term by $FV_F(M)$. The free (first-order) variables of a context $\g$
are denoted by $FV(\g)$ ($FV_F(\g)$) and defined in a natural way.

\subsection{The reduction relation}
The deterministic reduction relation $\to$ arises from the
following reduction rules and evaluation contexts:
\[
(\lambda x : \phi.\ M) N \to M[x:=N] \qquad (\lambda a.\ M) t \to M[a:=t]
\]
\[
\FST(<M, N>) \to M \qquad \SND(<M, N>) \to N 
\]
\[
\CASE(\INL(M), x : \phi.\ N, x : \psi.\ O) \to N[x:=M] \qquad \CASE(\INR(M), x :
\phi.\ N, x : \psi.\ O) \to O[x:=M]
\]
\[
\LET\ [a, x : \phi] := [t, M]\ \IN\ N \to N[a:=t][x:=M]
\]
\[
\qquad \pl{axProp}(t, \ov{u}, \pl{axRep}(t, \ov{u}, M)) \to M
\]
\[
\IND_{\phi}(M, \overline{t}) \to \lambda c.\ M\ c\
(\lambda b. \lambda x : b \ini c.\ \IND_{\phi}(M, \overline{t})\ b)
\]

In the reduction rules for $\IND$ terms, the variable $x$ is new.

The evaluation contexts describe call-by-need (lazy) evaluation order:
\[
[ \circ ] ::= \FST([ \circ ])\ |\ \SND([ \circ ])\ |\ \CASE([ \circ ], x. N,
x.O)\
\]
\[
\pl{axProp}(t, \ov{u}, [ \circ ])\ |\ \LET\ [a, x : \phi] := [ \circ ]\ \IN\ N\ |\
[ \circ ]\ M\ |\ \pl{magic}([\circ])
\]

We distinguish certain $\lii$ terms as values. The values are generated
by the following abstract grammar, where $M$ is an arbitrary term.
Obviously, there are no possible reductions from values.
\[
V ::= \lambda a.\ M\ |\ \lambda x : \phi.\ M\ |\ \INR(M)\ |\ \INL(M)\ |\ [t, M]\ |\ <M, N>\ |\ \pl{axRep}(t, \ov{u}, M)
\]

\begin{definition}
We write $M \downarrow$ if the reduction sequence starting from $M$
terminates. In this situation we also say that $M$ \emph{normalizes}. We write $M \downarrow v$ if we want to state that $v$
is the term at which this reduction sequence terminates. We write $M \to^*
M'$ if $M$ reduces to $M'$ in some number of steps. 
\end{definition}

\subsection{The types of $\lii$}\label{types}

The type system for $\lii$ is constructed according to the principle
of the Curry-Howard isomorphism for \izfio. Types are \izfio\ formulas, and terms are
$\li$ terms. Contexts $\g$ are finite sets of pairs $(x_i, \phi_i)$. The
first set of rules corresponds to first-order logic. 
\[
\infer{\g, x : \phi \p x : \phi}{} \qquad \infer{\gp M\ N : \psi}{\gp M : \phi \to
\psi & \gp N : \phi} \qquad \infer{\gp \lambda x : \phi.\ M : \phi \to
\psi}{\g, x : \phi \p M : \psi}
\]
\[
\infer{\gp <M, N> : \phi \land \psi}{\gp M : \phi & \gp N : \psi} \qquad
\infer{\gp \FST(M) : \phi}{\gp M : \phi \land \psi} \qquad \infer{\gp \SND(M) :
\psi}{\gp M : \phi \land \psi}
\]
\[
\infer{\gp \INL(M) : \phi \lor \psi}{\gp M : \phi} \qquad \infer{\gp \INR(M)
: \phi \lor \psi}{\gp M : \psi} 
\]
\[
\infer{\gp \CASE(M, x : \phi.\ N, x : \psi.\ O) : \vartheta}{\gp M : \phi \lor \psi & \g, x : \phi \proves N : \vartheta & \g, x : \psi \proves O : \vartheta}
\]
\[
\infer[a \notin FV_F(\g)]{\gp \lambda a.\ M : \forall a.\
\phi}{\g \proves M : \phi} \qquad \infer{\gp M\ t :
\phi[a:=t]}{\gp M : \forall a.\ \phi} \qquad 
\infer{\gp [t, M] : \exists a.\ \phi}{\gp M : \phi[a:=t]}
\]
\[
\infer{\gp \MAGIC(M) : \phi}{\gp M : \bot} \qquad
\infer[a \notin FV_F(\g, \psi)]{\gp \LET\ [a, x : \phi] := M\ \IN\
N : \psi}{\gp M : \exists a.\ \phi & \g,  x : \phi \proves N : \psi}
\]
The rest of the rules correspond to \izfio\ axioms:

\[
\infer{\gp \pl{eqRep}(t, u, M) : t \eqo u}{\gp M : \forall d.\ (d \ini t \to d
\ino u) \land (d \ini u \to d \ino t)}
\]
\[
\infer{\gp \pl{eqProp}(t, u, M) : \forall d.\ (d \ini t \to d
\ino u) \land (d \ini u \to d \ino t)}{\gp M : t \eqo u}
\]
\[
\infer{\gp \pl{inRep}(t, u, M) : t \ino u}{\gp M : \exists c.\ c \ini u \land
t \eqo c} \qquad 
\infer{\gp \pl{inProp}(t, u, M) : \exists c.\ c \ini u \land t \eqo c}{\gp t \in u}
\]
\[
\infer{\gp \pl{axRep}(t, \ov{u}, M) : t \ini t_A(\ov{u})}{\gp M : \phi_A(t,
\ov{u})} \qquad 
\infer{\gp \pl{axProp}(t, \ov{u}, M) : \phi_A(t, \ov{u}) }{ \gp M : t
\ini t_A(\ov{u})}
\]
\[
\infer{\gp \IND_{\phi(a, \ov{b})}(M, \ov{t}) : \forall a.\ \phi(a, \ov{t})}{\gp M : \forall c.\
(\forall b.\ b \ini c \to \phi(b, \ov{t})) \to \phi(c, \ov{t})}
\]

\subsection{The properties of $\lii$}

We now proceed with a standard sequence of lemmas for $\lii$.

\begin{lemma}[Canonical Forms]
Suppose $M$ is a value and $\proves M : \vartheta$. Then: 
\begin{enumerate}[$\bullet$]
\item $\vartheta = t \ini t_A(\ov{u})$ iff $M = \pl{axRep}(t, \ov{u}, N)$ and $\p N : \phi_A(t, \ov{u})$.
\item $\vartheta = \phi \lor \psi$ iff  ($M = \INL(N)$ and $\p N : \phi$) or ($M =
\INR(N)$ and $\p N : \psi$).
\item $\vartheta = \phi \land \psi$ iff $M = <N, O>$, $\p N : \phi$ and $\p
O : \psi$. 
\item $\vartheta = \phi \to \psi$ iff $M = \lambda x : \phi.\ N$ and $x :
\phi \p N : \psi$. 
\item $\vartheta = \forall a.\ \phi$ iff $M = \lambda a.\ N$ and $\p N :
\phi$.
\item $\vartheta = \exists a.\ \phi$ iff $M = [t, N]$ and $\p N : \phi[a:=t]$.
\item $\vartheta = \bot$ never happens.
\end{enumerate}
\end{lemma}
\begin{proof}
Immediate from the typing rules and the definition of values. 
\end{proof}

\begin{lemma}[Weakening]
If $\gp M : \phi$ and $FV(\psi) \cup \{ x \}$ are fresh to the proof tree $\gp M : \phi$, then $\g, x : \psi \p M : \phi$.
\end{lemma}
\begin{proof}
Straightforward induction on $\gp M : \phi$. 
\end{proof}

There are two substitution lemmas, one for the propositional part, the other
for the first-order part of the calculus. Since the rules and terms of $\lii$
corresponding to \izfio\ axioms do not interact with substitutions in a
significant way, the proofs are routine. 

\begin{lemma}\label{lamsl}
If $\Gamma, x : \phi \proves M : \psi$ and  $\Gamma \proves N : \phi$, then
$\Gamma \proves M[x:=N] : \psi$.
\end{lemma}
\proof 
By induction on $\g, x : \phi \p  M : \psi$. We show two interesting cases.
\begin{enumerate}[$\bullet$]
\item $\psi = \psi_1 \to \psi_2$, $M = \lambda y : \psi_1.\ O$. Using $\alpha$-conversion 
we can choose $y$ to be new, so that $y \notin FV(\g, x) \cup FV(N)$. The
proof tree must end with:
\[
\infer{\g, x : \phi \p \lambda y : \psi_1.\ O : \psi_1 \to \psi_2}{\g, x :
\phi, y : \psi_1 \p O : \psi_2}
\]
By the inductive hypothesis, $\g, y : \psi_1 \p O[x:=N] : \psi_2$, so $\g \p \lambda y : \psi_1.\ O[x:=N] : \psi_1
\to \psi_2$. By the choice of $y$, $\gp (\lambda y : \psi_1.\ O)[x:=N] :
\psi_1 \to \psi_2$. 
\item $\psi = \psi_2, M = \LET\ [a, y : \psi_1] := M_1\ \IN\ M_2$. The proof tree ends with:
\[
\infer{\g, x : \phi \p \LET\ [a, y : \psi_1] := M_1\ \IN\
M_2 : \psi_2}{\g, x : \phi \p M_1 : \exists a.\ \psi_1 & \g, x : \phi, y : \psi_1 \p M_2
: \psi_2}
\]
Choose $a$ and $y$ to be fresh. By the inductive hypothesis, $\gp M_1[x:=N] : \exists a.\ \psi_1$ and $\g, y
: \psi_1 \p M_2[x:=N] : \psi_2$. Thus $\gp \LET\ [a, y : \psi_1] :=
M_1[x:=N]\ \IN\ M_2[x:=N] : \psi_2$. By $a$ and $y$ fresh, $\gp (\LET\ [a, y : \psi_1] :=
M_1\ \IN\ M_2)[x:=N] : \psi_2$ which is what we want.\qed 
\end{enumerate}

\begin{lemma}\label{logsl}
If $\gp M : \phi$, then $\Gamma[a:=t] \proves M[a:=t] : \phi[a:=t]$.
\end{lemma}
\proof 
By induction on $\gp M : \phi$. Most of the rules do not interact with
first-order substitution, so we will show the proof just for two of them which
do. 
\begin{enumerate}[$\bullet$]
\item $\phi = \forall b.\ \phi_1$, $M = \lambda b.\ M_1$. The proof tree ends with:
\[
\infer[b \notin FV_F(\g)]{\gp \lambda b.\ M_1 : \forall b.\ \phi_1}{\g \proves M_1 : \phi_1}
\]
Without loss of generality we can assume that $b \notin FV(t) \cup \{ a \}$. By the inductive hypothesis, $\Gamma[a:=t] \proves M_1[a:=t] :
\phi_1[a:=t]$. Therefore $\g[a:=t] \p \lambda b.\ M_1[a:=t] : \forall
b.\ \phi_1[a:=t]$ and by the choice of $b$, $\g[a:=t] \p (\lambda b.\ M_1)[a:=t] \p (\forall b.\ \phi_1)[a:=t]$. 
\item $\phi = \phi_1[b:=u]$, $M = M_1\ u$ for some term $u$. The proof tree ends with:
\[
\infer{\gp M_1\ u : \phi_1[b:=u]}{\gp M_1 : \forall b.\ \phi_1}
\]
Choosing $b$ to be fresh, by the inductive hypothesis we get $\g[a:=t] \p
M_1[a:=t] : \forall b.\ (\phi_1[a:=t])$, so $\g[a:=t] \p M_1[a:=t]\ u[a:=t] :
\phi_1[a:=t][b:=u[a:=t]]$. By Lemma \ref{formsubst} and $b \notin FV(t)$, we
get $\g[a:=t] \p (M_1\ u)[a:=t] : \phi_1[b:=u][a:=t]$.\qed
\end{enumerate}

With the lemmas at hand, Progress and Preservation follow easily:

\begin{lemma}[Subject Reduction, Preservation]
If $\gp M : \phi$ and $M \to N$, then $\gp N : \phi$.
\end{lemma}
\begin{proof}
By induction on the definition of $M \to N$. We show several cases. Case $M
\to N$ of:
\begin{enumerate}[$\bullet$]
\item $(\lambda x : \phi_1.\ M_1)\ M_2 \to M_1[x:=M_2]$. The proof tree $\gp M
: \phi$ must end with:
\[
\infer{\gp (\lambda x : \phi_1.\ M_1)\ M_2 : \phi}
{
  \infer
  {
    \gp \lambda x : \phi_1.\ M_1 : \phi_1 \to \phi
  }
  {
    \g, x : \phi_1 \p M_1 : \phi
  }
  & 
  \gp M_2 : \phi_1
}
\]
By Lemma \ref{lamsl}, $\gp M_1[x:=M_2] : \phi_1$.
\item $\LET\ [a, x : \phi_1] := [t, M_1]\ \IN\ M_2 \to M_2[a:=t][x:=M_1]$. The
proof tree $\gp M : \phi$ must end with:
\[
\infer
{\gp \LET\ [a, x : \phi_1] := [t, M_1]\ \IN\ M_2 : \phi}
{
  \infer{\gp [t, M_1] : \exists a.\ \phi_1}
  {\gp M_1 : \phi_1[a:=t]}
  & 
  \g, x : \phi_1 \proves M_2 : \phi
}
\]
Choose $a$ to be fresh. Thus $M_1[a:=t] = M_1$ and $\g[a:=t] = \g$. By the side-condition of the last
typing rule, $a \notin FV(\phi)$, so $\phi[a:=t] = \phi$. By Lemma
\ref{logsl} we get $\g[a:=t], x : \phi_1[a:=t] \p M_2[a:=t] : \phi[a:=t]$,
so also $\g, x : \phi_1[a:=t] \p M_2[a:=t] : \phi$. By Lemma \ref{lamsl}, we
get $\gp M_2[a:=t][x:=M_1] : \phi$.
\item $\pl{axProp}(t, \ov{u}, \pl{axRep}(t, \ov{u}, M_1)) \to M_1$.
The proof tree must end with:
\[
\infer{\gp \pl{axProp}(t, \ov{u}, \pl{axRep}(t, \ov{u}, M_1)) : \phi_A(t, \ov{u})}
{
\infer{\gp \pl{axRep}(t, \ov{u}, M_1)) : t \ini t_A(\ov{u})}{\gp M_1 :
\phi_A(t, \ov{u})}
}
\]
The claim follows immediately.
\item $\IND_{\psi(a, \ov{f})}(M_1, \ov{t}) \to \lambda c.\ M_1\ c\
(\lambda b. \lambda x : b \ini c.\ \IND_{\psi(a, \ov{b})}(M_1, \ov{t})\ b)$. The proof tree must end with:
\[
\infer{\gp \IND_{\psi(a, \ov{f})}(M_1, \ov{t}) : \forall a.\ \psi(a,
\ov{t})}{\gp M_1 : \forall c.\ (\forall b.\ b \ini c \to \psi(b, \ov{t})) \to \psi(c, \ov{t})}
\]
We choose $b, c, x$ to be fresh. By applying $\alpha$-conversion we can also obtain a proof
tree of $\gp M_1 : \forall e.\ (\forall d.\ d \ini e \to \psi(d, \ov{t}))
\to \psi(e, \ov{t})$, where $\{ d, e \} \cap \{ b, c \} = \emptyset$. Then
by Weakening we get $\g, x : b \ini c \p M_1 : \forall e.\ (\forall d.\ d
\ini e \to \psi(d, \ov{t})) \to \psi(e, \ov{t})$, so also $\g, x : b \ini c \p \IND_{\psi(a, \ov{b})}(M_1, \ov{t})
: \forall a.\ \psi(a, \ov{t})$. Let the proof tree $T$ be defined as:
\[
    \infer{\gp \lambda b. \lambda x : b \ini c.\ \IND_{\psi(a,
\ov{b})}(M_1, \ov{t})\ b : \forall b.\ b \ini c \to \psi(b,
\ov{t})}
    {
      \infer{\gp \lambda x : b \ini c.\ \IND_{\psi(a,
\ov{b})}(M_1, \ov{t})\ b : b \ini c \to \psi(b,
\ov{t})}
      {
        \infer{\g, x : b \ini c \p \IND_{\psi(a, \ov{b})}(M_1,
\ov{t})\ b : \psi(b, \ov{t})}
        {
	  \g, x : b \ini c \p \IND_{\psi(a, \ov{b})}(M_1, \ov{t})
: \forall a.\ \psi(a, \ov{t})
	}
      }
    }
\]
Then the following proof tree shows the claim:
\[
\infer{\gp \lambda c.\ M_1\ c\ (\lambda b. \lambda x : b \ini c.\ \IND_{\psi(a,
\ov{b})}(M_1, \ov{t})\ b) : \forall c.\ \psi(c, \ov{t})}
{
  \infer
  {
    \gp M_1\ c\ (\lambda b. \lambda x : b \ini c.\ \IND_{\psi(a,
\ov{b})}(M_1, \ov{t})\ b) : \psi(c, \ov{t})
  }
  {
    \infer{\gp M_1\ c : (\forall b.\ b \ini c \to \psi(b, \ov{t})) \to \psi(c, \ov{t})}
    {
      \gp M_1 : \forall c.\ (\forall b.\ b \ini c \to \psi(b, \ov{t})) \to
\psi(c, \ov{t})
    }
    &
    T
  }
}
\]
\end{enumerate}
\end{proof}
\begin{lemma}[Progress]
If $\ \proves M : \phi$, then either $M$ is a value or there is $N$ such that $M \to N$.
\end{lemma}
\proof 
Straightforward induction on the length of $M$. The proof proceeds by case analysis of $M$. We show several cases:
\begin{enumerate}[$\bullet$]
\item It is easy to see that the case $M = x$ cannot happen.
\item If $M = \lambda x : \phi.\ N$, then $M$ is a value.
\item If $M = N\ O$, then for some $\psi$, the proof must end with:
\[
\infer{\p N\ O : \phi}{\p N : \psi \to \phi & \p O : \psi}
\]
By the inductive hypothesis, either $N$ is a value or there is $N'$ such
that $N \to N'$. In the former case, by Canonical Forms for some $P$ we have $N = \lambda x :
\psi.\ P$, so $N\ O \to P[x:=O]$. In the latter case, $N\ O \to N'\ O$. 
\item If $M = \pl{axRep}(t, \ov{u}, M)$, then $M$ is a value.
\item If $M = \pl{axProp}(t, \ov{u}, O)$, then we have the following proof tree:
\[
\infer{\p \pl{axProp}(t, \ov{u}, O) : \phi_A(t, \ov{u})}
{
\p O : t \ini t_A(\ov{u})
}
\]
By the inductive hypothesis, either $O$ is a value or there is $O_1$ such
that $O \to O_1$. In the former case, by Canonical Forms, $O = \pl{axRep}(t,
\ov{u}, P)$ and $M \to P$. In the latter, by the evaluation rules $\pl{axProp}(t,
\ov{u}, O) \to \pl{axProp}(t, \ov{u}, O_1)$.
\item The cases corresponding to the equality and membership axioms work in the same way.
\item The $\pl{ind}$ terms always reduce.\qed
\end{enumerate}

\begin{corollary}\label{corlz}
If $\ \p M : \phi$ and $M \downarrow v$, then $\p v : \phi$ and $v$ is a value.
\end{corollary}

\begin{corollary}\label{corbot}
If $\p M : \bot$, then $M$ does not normalize.
\end{corollary}
\begin{proof}
If $M$ normalized, then by Corollary \ref{corlz} we would have a value of
type $\bot$, which by Canonical Forms is impossible. 
\end{proof}

Finally, we state the formal correspondence between $\lii$ and \izfio:

\begin{lemma}[Curry-Howard isomorphism]
If $\gp  O : \phi$ then \izfio $ + rg(\g) \p  \phi$, where $rg(\g) = \{
\phi\ |\ (x, \phi) \in \g \}$. If \izfio $+ \g \p \phi$, then there exists a term $M$ such that $\og \p M :
\phi$, where $\og = \{ (x_\phi, \phi)\ |\ \phi \in \g \}$.
\end{lemma}
\begin{proof}
Both parts follow by easy induction on the proof. The first part is
straightforward, to get the claim simply erase the lambda terms from the
proof tree. For the second part, we show terms and trees corresponding to \izfio\ axioms:
\begin{enumerate}[$\bullet$]
\item Let $\phi$ be one of the \izfio\ axioms apart from $\in$-Induction.
Then $\phi = \forall \ov{a}.\ \forall c.\ c \ini t_A(\ov{a}) \iffl \phi_A(c,
\ov{a})$ for the axiom (A) (incorporating axioms (IN) and (EQ) in this case
in the obvious way). Recall that $\phi_1 \iffl \phi_2$ is an
abbreviation for $(\phi_1 \to \phi_2) \land (\phi_2 \to \phi_1)$. Let $T$
be the following proof tree:
\[
    \infer{\gp \lambda x : \phi_A(c, \ov{a}).\ \pl{axRep}(c, \ov{a}, x) : \phi_A(c, \ov{a}) \to c \ini t_A(\ov{a})}
    {
      \infer{\g, x : \phi_A(c, \ov{a}) \p \pl{axRep}(c, \ov{a}, x) : c \ini t_A(\ov{a})}
      {
        \g, x : \phi_A(c, \ov{a}) \p x : \phi_A(c, \ov{a})
      }
    }
\]
Let $M_1 = \lambda x : c \ini t_A(\ov{a}).\ \pl{axProp}(c, \ov{a}, x)$ and
let $M_2 = \lambda x : \phi_A(c, \ov{a}).\ \pl{axRep}(c, \ov{a}, x)$. 
Then the following proof tree shows the claim:
\[
\infer{\gp \lambda \ov{a} \lambda c.\ <M_1,M_2>
  : \forall
\ov{a}.\ \forall c.\ c \ini t_A(\ov{a}) \iffl \phi_A(c, \ov{a})}
{
  \infer
  {\gp <M_1, M_2> : c \ini t_A(\ov{a}) \iffl \phi_A(c, \ov{a})}
  {
    \infer
    {
    \gp M_1 : c \ini t_A(\ov{a}) \to \phi_A(c, \ov{a})
    }
    {
      \infer{\g, x : c \ini t_A(\ov{a}) \p \pl{axProp}(c, \ov{a}, x) : \phi_A(c, \ov{a})}
      {
       \g, x : c \ini t_A(\ov{a}) \p x : c \ini t_A(\ov{a})
      }
    }
    &
    \qquad
    T
  }
}
\]
\item Let $\phi$ be the $\in$-induction axiom. Let
\[
M = \lambda \ov{f} \lambda x : (\forall a.
(\forall b.\ b \ini a \to \psi(b, \ov{f})) \to \psi(a, \ov{f})).\ \IND(x, \ov{f}).
\]
The following proof tree shows the claim:
\[
\infer{\gp M : \forall \ov{f}. (\forall a. (\forall b.\ b \ini a \to
\psi(b, \ov{f})) \to \psi(a, \ov{f})) \to \forall a.\ \psi(a, \ov{f})}
{
  \infer{\g, x : \forall a. (\forall b.\ b \ini a \to \phi(b, \ov{f})) \to
\psi(a, \ov{f}) \p \IND_{\psi(a, \ov{f})}(x, \ov{f}) : \forall a.\ \psi(a, \ov{f})}
  {
    \g, x : \forall a. (\forall b.\ b \ini a \to \psi(b, \ov{f})) \to
\psi(a, \ov{f}) \p x : \forall a. (\forall b.\ b \ini a \to \psi(b, \ov{f})) \to
\psi(a, \ov{f})
  }
}
\]
\end{enumerate}
\end{proof}

Note that all proofs in this section are constructive and quite weak from
the proof-theoretic point of view --- Heyting Arithmetic should be
sufficient to formalize the arguments. However, by the Curry-Howard isomorphism
and Corollary \ref{corbot}, normalization of $\lii$ entails consistency of \izfio,
which easily interprets Heyting Arithmetic. Therefore a normalization
proof must utilize much stronger means, which we introduce in the following
section. 

\newcommand\vl{V^{\lambda}}
\newcommand\vla{\vl_\alpha}
\newcommand\vlb{\vl_\beta}
\newcommand\lovz{\Lambda_{\overline{Z}}}

\section{Realizability for \izfio}\label{izfreal}

In this section we work in ZF with $\omega$-many strongly inaccessible
cardinals. We denote the $i$-th strongly inaccessible by \inac{i} and choose
them so that $\inac{i} \in \inac{i+1}$. It is likely that IZF with
Collection and $\omega$-many inaccessible sets would be sufficient, as
excluded middle is not used explicitly; however, arguments using ordinals
and ranks would need to be done very carefully, as the notion of an ordinal
in constructive set theories is problematic \cite{powell, taylor96}.

\subsection{Realizers} 

Our realizers are essentially terms of $\lii$. For convenience, wherever
possible, we erase logic terms and formulas from parameters of $\pl{axRep},
\pl{axProp}$, $\pl{ind}$ and $\pl{case}$ terms. We call the resulting calculus $\la$. More
formally, $\la$ arises as an image of an erasure map $\overline{M}$, which takes
as its argument a $\lii$-term. This map is defined by structural induction on
$M$ and induced by the following cases:
\[
\overline{\pl{axRep}(t, \ov{u}, M)} = \pl{axRep}(\overline{M}) \qquad \overline{\pl{axProp}(t,
\ov{u}, M)} = \pl{axProp}(\overline{M}) \qquad \overline{\IND_\phi(M, \ov{t})} = \IND(\overline{M})
\]
\[
\overline{\lambda x : \phi.\ M} = \lambda x.\ \overline{M} \qquad \overline{\LET\ [a, x :
\phi] :=M\ \IN \ N} = \LET\ [a, x] :=\overline{M}\ \IN \ \overline{N}
\]
\[
\overline{\CASE(M, x : \phi.\ N, x : \psi.\ O)} = \CASE(\overline{M}, x.\overline{N}, x.\overline{O})
\]
The erasure on the rest of terms is defined in a natural way, for example
$\overline{<M, N>} = <\overline{M}, \overline{N}>$, $\overline{[t, M]} = [t, \overline{M}]$ and $\overline{M\ t}
= \overline{M}\ t$. The reduction rules and values in $\la$ are induced from $\lii$ in an
obvious way. The set of $\la$ terms will be denoted by
$\lat$ and the set of $\la$ values will be denoted by $\la_v$.

\begin{lemma}\label{erasurenorm}
If $\overline{M}$ normalizes, so does $M$. 
\end{lemma}
\begin{proof}
Straightforward --- the erased information does not affect the reductions. 
\end{proof}

The fact that logic terms do not play any role in the reductions is crucial
for the normalization argument to work. 

This definition of the erasure map and $\la$ fixes a small mistake in the
presentation in \cite{jacsl2006}, where a bit too much information was erased.

\subsection{Realizability relation}

Having defined realizers, we proceed to define the realizability relation.
Our definition was inspired by McCarty's \cite{mccarty}. From now on, the
letter $T$ denotes the set of all \izfio\ terms. 

\begin{definition}
A set $A$ is a $\lambda$-name iff $A$ is a set of pairs $(v, B)$ such that
$v \in \la_v$ and $B$ is a $\lambda$-name.
\end{definition}

In other words, $\lambda$-names are sets hereditarily labelled by $\la$ values.

\begin{definition}
The class of $\lambda$-names is denoted by $\vl$.
\end{definition}

Formally, $\vl$ is generated by the following transfinite inductive
definition on ordinals:
\[
V^\lambda_\alpha = \bigcup_{\beta \lt \alpha} P(\la_v \times
V^\lambda_\beta) \qquad V^\lambda = \bigcup_{\alpha \in \mbox{ORD}}V^\lambda_\alpha
\]

\begin{definition}
The \emph{$\lambda$-rank} of a $\lambda$-name $A$, denoted by $\lrk(A)$, is the
smallest $\alpha$ such that $A \in \vla$.  
\end{definition}

We now define three auxiliary relations between $\la$ terms and pairs of
sets in $\vl$, which we write as $M \reals A \ini B$, $M \reals A \in B$, $M
\reals A = B$. These relations are a prelude to the definition of realizability. 

\[
\begin{array}{lcl}
M \reals A \ini B & \equiv & M \downarrow v \land (v, A) \in B \\
M \reals A \in B & \equiv & M \downarrow \pl{inRep}(N) \land N \downarrow [u, O] \land \exists C \in \vl.\ O \downarrow <O_1, O_2> \land \\
& & O_1 \reals C \ini B \land O_2 \reals A = C \\
M \reals A \eqo B & \equiv & M \downarrow \pl{eqRep}(M_0) \land M_0
\downarrow \lambda a.\ M_1 \land \forall t \in T, \forall D \in \vl.\ M_1[a:=t] \downarrow <O, P> \land\\
& & O \downarrow \lambda x.\ O_1 \land \forall N.\ (N \reals D \ini A) \to O_1[x:=N] \reals D \in B \land\\
& & P \downarrow \lambda x.\ P_1 \land \forall N.\ (N \reals D \ini B) \to P_1[x:=N] \reals D \in A
\end{array}
\]

The relations $M \reals A \in B$ and $M \reals A = B$ are defined together
in a standard way by transfinite recursion. See for example \cite{rathjendp} for more details.

\begin{definition}
For any set $C \in \vl$, $\rin{C}$ denotes $\{ (M, A)\ |\ M \reals A \ino C \}$.
\end{definition}

\begin{definition}
A (class-sized) first-order language $L$ arises from enriching the \izfio\ signature
with constants for all $\lambda$-names.
\end{definition}

From now on until the end of this section, symbols $M, N, O, P$ range exclusively over $\la$-terms, letters $a, b, c$ vary over 
first-order variables in the language, letters $A, B, C$ vary over $\lambda$-names
and letter $\rho$ varies over finite partial functions from first-order variables
in $L$ to $V^\lambda$. We call such functions \emph{environments}.

\begin{definition}
For any formula $\phi$ of $L$, any term $t$ of $L$ and $\rho$ defined on all free variables of
$\phi$ and $t$, we define by metalevel induction a realizability relation $M
\reals_\rho \phi$ in an environment $\rho$ and a meaning of a term  
$\SB{t}_\rho$ in an environment $\rho$:
\begin{enumerate}[(1)]
\item $\SB{a}_\rho \equiv \rho(a)$
\item $\SB{A}_\rho \equiv A$
\item \label{omegadef} $\SB{\omega}_\rho \equiv \omega'$, where $\omega'$ is defined by the means
of inductive definition: $\omega'$ is the smallest set such that:
\begin{enumerate}[$\bullet$]
\item $(\pl{infRep}(N), A) \in \omega'$ if $N \downarrow \INL(O)$, $O \rrho A =
0$ and $A \in \vl_\omega$.
\item If $(M, B) \in \omega'^+$, then $(\pl{infRep}(N), A) \in \omega'$ if $N
\downarrow \INR(N_1)$, $N_1 \downarrow [t, O]$, $O \downarrow <M, P>$, $P
\rrho A = S(B)$ and $A \in \vl_\omega$.
\end{enumerate}
Note that if $(M, B) \in \omega'^+$, then there is a finite ordinal $\alpha$
such that $B \in \vl_\alpha$. 
\item \label{inacdef} $\sr{V_i} \equiv \vis{i}$. We will define $\vis{i}$ below.
\item \label{termdef} $\SB{t_A(\ov{u})}_\rho \equiv \{ (\pl{axRep}(N),B) \in
\la_v \times \vl_\gamma\ |\ N \reals_\rho \phi_A(B,
\overrightarrow{\SB{u}_\rho)}\}$.
The ordinal $\gamma$ will be defined below. 
\item $M \reals_\rho \bot \equiv \bot$
\item $M \reals_\rho t \ini s \equiv M \reals \sr{t} \ini \sr{s}$
\item $M \reals_\rho t \in s \equiv M \reals \sr{t} \in \sr{s}$
\item $M \rrho t \eqo s \equiv M \reals \sr{t} = \sr{s}$
\item $M \reals_\rho \phi \land \psi \equiv M \downarrow <M_1, M_2> \land (M_1
\reals_\rho \phi) \land (M_2 \reals_\rho \psi)$
\item $M \reals_\rho \phi \lor \psi \equiv (M \downarrow \INL(M_1) \land M_1
\reals_\rho \phi) \lor (M \downarrow \INR(M_1) \land M_1 \reals_\rho \psi)$
\item $M \reals_\rho \phi \to \psi \equiv (M \downarrow \lambda x.\ M_1) \land
\forall N.\ (N \reals_\rho \phi) \to (M_1[x:=N] \reals_\rho \psi)$
\item $M \reals_\rho \exists a.\ \phi \equiv M \downarrow [t, N] \land \exists A \in
V^\lambda.\ N \reals_\rho \phi[a:=A]$
\item $M \reals_\rho \forall a.\ \phi \equiv M \downarrow \lambda a.\ N \land \forall A \in
V^\lambda, \forall t \in T.\ N[a:=t] \reals_\rho \phi[a:=A]$
\end{enumerate}
\end{definition}

To define \vis{i}, first recall that the axiom (INAC${}_i$) has the
following form:
\[
(\mbox{INAC}_i)\ \forall c.\ c \in V_i \iffl \phiinaca(c, V_i) \land
\forall d.\ \phiinacb(d) \to c \in d.
\]
We define a monotonic operator $F$ on sets as:
\[
F(A) = A \cup \{ (\pl{inac}_i\pl{Rep}(N), C) \in \la_v \times
\vinac{i}\ |\ N \rrho \phiinaca(C, A) \land \forall d.\ \phiinacb(d) \to C
\in d \}.
\]
We set \vis{i} to be the smallest fixpoint of $F$. Formally, \vis{i} is generated by transfinite inductive definition on ordinals:
\[
\vis{i, \gamma} = F(\bigcup_{\beta \lt \gamma} \vis{i, \beta})
\qquad \vis{i} = \bigcup_{\gamma \in \mbox{ORD}} \vis{i, \gamma}
\]
Since $F$ adds only elements from $\la_v \times \vinac{i}$, any element
of $\vis{i}$ is in $\la_v \times \vinac{i}$, so $\vis{i} \in
\vinac{i+1}$. 

The definition of the ordinal $\gamma$ in item \ref{termdef} 
depends on $t_A(\ov{u})$. This ordinal is close to the rank of the set denoted
by $t_A(\ov{u})$ and is chosen so that Lemma \ref{realsterms} can be proved.
Let $\ov{\alpha} =  \overrightarrow{\lrk(\SB{u}_\rho)}$. Case $t_A(\ov{u})$ of:
\begin{enumerate}[$\bullet$]
\item $\{ u_1, u_2 \}$ --- $\gamma = max(\alpha_1, \alpha_2)$
\item $P(u)$ --- $\gamma = \alpha + 1$. 
\item $\bigcup u$ --- $\gamma = \alpha$.
\item $S_{\phi(a, \ov{f})}(u, \ov{u})$ --- $\gamma = \alpha_1$.
\item $R_{\phi(a, b, \ov{f})}(u, \ov{u})$. This case is more complicated.
The names are chosen to match the corresponding clause in the proof of Lemma \ref{realsterms}. 
Let $G = \{ (N_1, (N_{21}, B)) \in \lat \times \SB{u}^+_\rho\ |\
\exists d \in \vl.\ \psi(N_1, N_{21}, B, d) \}$, where
$\psi(N_1, N_{21}, B, d) \equiv (N_1 \downarrow \lambda a.\ N_{11}) \land (N_{11}
\downarrow \lambda x.\ O) \land (O[x:=N_{21}] \reals_\rho
\phi(B, d, \overrightarrow{\SB{u}_\rho}) \land \forall e.\ \phi(B, e,
\overrightarrow{\SB{u}_\rho}) \to
e = d)$. Then for all $g \in G$ there is $D$ and $(N_1, (N_{21}, B))$ such that $g =
(N_1, (N_{21}, B))$ and $\psi(N_1, N_{21}, B, D)$. Use Collection to collect these $D$'s in one set $H$, so that for
all $g \in G$ there is $D \in H$ such that the property holds. Apply Replacement
to $H$ to get the set of $\lambda$-ranks of sets in $H$. Then $\beta \equiv \bigcup H$ is
an ordinal and for any $D \in H$, $\lrk(D) \lt \beta$. Therefore for all $g \in G$ there is $D \in \vl_\beta$ and $(N_1, (N_{21}, B))$ such that $g =
(N_1, (N_{21}, B))$ and $\psi(N_1, N_{21}, B, D)$ holds. Set $\gamma = \beta + 1$.
\end{enumerate}

At this point it is not clear yet that the realizability definition makes sense
--- a priori it might be circular. We will now show that it is not the case. 

\begin{definition}
For any closed term $s$, we define number of occurences of $s$ in any term $t$ and
formula $\phi$, denoted by $Occ(s, t)$ and $Occ(s, \phi)$, respectively, by induction on the definition of terms and formulas. We
show representative clauses of the definition:
\begin{enumerate}[$\bullet$]
\item $Occ(s, s) = 1$. 
\item $Occ(s, a) = 0$, where $a$ is a variable.
\item $Occ(s, t_A(\ov{u})) = Occ(s, u_1) + {\ldots} + Occ(s, u_n)$. 
\item $Occ(s, S_{\phi}(t, \ov{u})) = Occ(s, \phi) + Occ(s, t) + Occ(s, u_1)
+ {\ldots}  + Occ(s, u_n)$. 
\item $Occ(s, t \in u) = Occ(s, t) + Occ(s, u)$.
\item $Occ(s, \phi \land \psi) = Occ(s, \phi) + Occ(s, \psi)$. 
\end{enumerate}
In a similar manner we define the number of function symbols $FS$ in a
term and formula. 
\end{definition}

\begin{definition}
Let $M(\nat)$ denote the set of all multisets over $\nat$ with the
standard well-founded ordering. Formally, a member $A$ of $M(\nat)$ is a
function from $\nat$ to $\nat$, returning for any $n$ the number of copies
of $n$ in $A$. We define a function $V$ taking terms and
formulas into $M(\nat)$: $V(x)$ for any number $i$ returns $Occ(V_i,
x)$, for $x$ being either a term or a formula. 
\end{definition}

\begin{lemma}
The definition of realizability is well-founded. 
\end{lemma}
\begin{proof}
Use the measure function $m$ which takes a clause in the definition and
returns an element of $M(\nat) \times \nat^3$ with the lexicographical order:
\begin{eqnarray*}
m(M \reals_\rho \phi) &  = & (V(\phi), Occ(\omega, \phi), FS(\phi),
\mbox{``structural complexity of $\phi$''})\\
m(\SB{t}_\rho) & = & (V(t), Occ(\omega, t), FS(t), 0)
\end{eqnarray*}
Then the measure of the definiendum is always greater than the measure of
the definiens --- in the clauses for formulas the structural complexity goes
down, while the rest of parameters do not grow larger. In the definition of
$\sr{V_i}$, one $V_i$ disappears replaced by two $V_{i-1}$'s. In the definition
of $\sr{\omega}$, one $\omega$ disappears. Finally, in the definition of
$\sr{t_A(\ov{u})}$, the topmost $t_A$ disappears, while no new $V_i$'s and
$\omega$'s appear. 
\end{proof}

Since the definition is well-founded, (metalevel) inductive proofs on the
definition of realizability are justified, such as the proof of the following lemma:

\begin{lemma}\label{realsubst}
$\SB{t[a:=s]}_\rho = \SB{t[a:=\SB{s}_\rho]}_\rho = \SB{t}_{\rho[a:=\SB{s}_\rho]}$ and $M \reals_\rho
\phi[a:=s]$ iff $M \reals_\rho \phi[a:=\SB{s}_\rho]$ iff $M \reals_{\rho[a:=\SB{s}_\rho]} \phi$.
\end{lemma}
\begin{proof}
By induction on the definition of realizability. We show representative
cases. Case $t$ of:
\begin{enumerate}[$\bullet$]
\item $A$ --- then $\SB{t[a:=s]}_\rho = \SB{t[a:=\SB{s}_\rho]}_\rho =
\SB{t}_{\rho[a:=\SB{s}_\rho]} = A$. 
\item $a$ --- then $\sr{t[a:=s]} = \sr{s}$, $\sr{t[a:=\sr{s}]} = \sr{\sr{s}}
= \sr{s}$ and also $\SB{t}_{\rho[a:=\sr{s}]} = \sr{s}$.
\item $t_A(\ov{u})$. Then $\sr{t[a:=s]} = \{ (\pl{axRep}(N), A)\ |\ N \rrho
\phi_A(A, \ov{u}[a:=s]) \}$. By the inductive hypothesis, this is equal to $\{ (\pl{axRep}(N), A)\ |\ N
\reals_{\rho[a:=\sr{s}]} \phi_A(A, \ov{u}) \} =
\SB{t}_{\rho[a:=\sr{s}]}$ and also to $\{ (\pl{axRep}(N), A)\ |\ N \rrho
\phi_A(A, \ov{u}[a:=\sr{s}]) \}$ and thus to $\sr{t[a:=\sr{s}]}$.
\end{enumerate}
For formulas, the atomic cases follow by the proof above and the
non-atomic cases follow immediately by the application of the inductive
hypothesis. 
\end{proof}
\begin{lemma}\label{realnorm}
If $(M \rrho \phi)$ then $M \downarrow$.
\end{lemma}
\begin{proof}
Straightforward from the definition of realizability --- in every case the
definition starts with the clause assuring normalization of $M$. 
\end{proof}
\begin{lemma}\label{realredclosed}
If $M \to^* M'$ then $M'\reals_\rho \phi$ iff $M \reals_\rho \phi$.
\end{lemma}
\begin{proof}
Whether $M \rrho \phi$ or not depends only on the value of $M$, which does not
change with reduction or expansion.
\end{proof}

\begin{lemma}\label{afvreal}
If $\rho$ agrees with $\rho'$ on $FV(\phi)$, then $M \rrho \phi$ iff $M
\reals_{\rho'} \phi$. In particular, if $a \notin FV(\phi)$, then $M \rrho
\phi$ iff $M \reals_{\rho[a:=A]} \phi$. 
\end{lemma}
\begin{proof}
Straightforward induction on the definition of realizability --- the environment
is used only to provide the meaning of the free variables of terms in a
formula.
\end{proof}

\begin{lemma}\label{realimpl}
If $M \reals_\rho \phi \to \psi$ and $N \reals_\rho \phi$, then $M\ N \reals
\psi$. 
\end{lemma}
\begin{proof}
Suppose $M \reals_\rho \phi \to \psi$. Then $M \downarrow (\lambda x.\ O)$
and for all $P \reals \phi$, $O[x:=P] \reals \psi$. Now, $M\ N \to^*
(\lambda x.\ O)\ N \to O[x:=N]$. Lemma \ref{realredclosed} gives us the claim. 
\end{proof}

\subsection{Properties of realizability}

We now establish several properties of the realizability relation, which
mostly state that the truth in the realizability universe is not far from the truth in
the real world, as far as ranks of sets are concerned.

Several lemmas mirror similar facts from McCarty's thesis \cite{mccarty}. We
cannot, however, simply point to these lemmas and say that essentially they
prove the same thing, as our realizability behaves a bit differently from
his.

\begin{lemma}\label{ineqrank}
If $A \in \vla$, then there is $\beta \lt \alpha$ such that
for all $B$, if $M \reals_\rho B \in A$, then $B \in \vlb$. If $M
\reals_\rho B = A$, then $B \in \vla$. If $M \rrho B \ini A$, then $\lrk(B)
\lt \lrk(A)$. 
\end{lemma}
\begin{proof}
By induction on $\alpha$. Take any $A \in \vla$. By the definition of
$\vla$, there is $\beta \lt \alpha$ such that $A \subseteq \la_v \times
\vl_\beta$. Suppose $M \rrho B \in A$. Then $M \downarrow \pl{inRep}(N)$, $N
\downarrow [u, O]$, $O \downarrow <O_1, O_2>$ and there is $C$ such that
$O_1 \reals C \ini A$ and $O_2 \reals B = C$. Therefore, $O_1 \downarrow v$
and $(v, C) \in A$. Thus $C \in \vl_\beta$, so by the inductive hypothesis
also $B \in \vl_\beta$ and we get the claim of the first part of the lemma.

For the second part, suppose $M \rrho B = A$. This means that $M \downarrow
\pl{eqRep}(M_0)$, $M_0 \downarrow \lambda a.\ M_1$ and for all $t \in T, D$,
$M_1[a:=t] \downarrow <O, P>$. Moreover, $O \downarrow \lambda x.\ O_1$ and for all $N
\rrho D \ini B$ we have $O_1[x:=N] \rrho D \in A$. In particular, if $(v, D) \in B$, then $O_1[x:=v]
\rrho D \in A$. By the first part of the lemma, any such $D$ is
in $\vl_\beta$ for some $\beta \lt \alpha$, so $B \in \vla$.

The third part is trivial. 
\end{proof}

\begin{lemma}\label{eqin}
$M \rrho A \eqo B$ iff $M \downarrow \pl{eqRep}(N)$ and $N \rrho \forall d.\
(d \ini A \to d \ino B) \land (d \ini B \to d \ino A)$. Also, $M \rrho A
\ino B$ iff $M \downarrow \pl{inRep}(N)$ and $N \rrho \exists c.\ c \ini B
\land A = c$.
\end{lemma}
\begin{proof}
Simply expand what it means for $M$ to realize respective formulas. 
\end{proof}

We now exhibit realizers corresponding to proofs of Lemmas
\ref{eqrefl}-\ref{lei0}. Their existence and corresponding properties will follow immediately from
Theorem \ref{norm} once it is proved; however, we need them for the proof of Lemma
\ref{vfunclosed}. Since Lemma \ref{vfunclosed} only needs to be used
for a set theory with inaccessibles, an alternative to tedious proofs below
could be to prove normalization for the theory without inaccessibles first,
and take realizers from that normalization theorem. 

\begin{lemma}\label{realeqrefl}
There is a term $\pl{eqRefl}$ such that $\pl{eqRefl} \rrho
\forall a.\ a = a$. 
\end{lemma}
\begin{proof}
Take the term $\pl{eqRefl} \equiv \pl{ind}(M)$, where $M = \lambda c.\
\lambda x.\ \pl{eqRep}(\lambda d.\ <N, N>)$ and $N = \lambda y.\ \pl{inRep}([d, <y, x\ d\ y>])$. Then $\pl{eqRefl} \to \lambda a.\ M\ a\
(\lambda e.\ \lambda z.\ \pl{ind}(M)\ e)$. It suffices to
show that for any $A, t$, $M\ t\ (\lambda e.\ \lambda z.\
\pl{ind}(M)\ e) \rrho A = A$. We proceed by induction on
$\lambda$-rank of $A$. We have $M\ t\ (\lambda e.\ \lambda z.\ \pl{ind}(M)\ e) \downarrow \pl{eqRep}(\lambda d.\ <N, N>[x:=\lambda
e.\ \lambda z.\ \pl{ind}(M)\ e])$. It suffices to
show that for all $s \in T, D \in \vl$, for all $O \rrho D \ini A$,
$\pl{inRep}([s, [O, (\lambda e.\ \lambda z.\ \pl{ind}(M)\ e)\ s\ O>]) \rrho D \in A$.
Take any $s, D$ and $O \rrho D \ini A$. By Lemma \ref{ineqrank}, $\lrk(D) \lt \lrk(A)$. 
We need to show the existence of $C$ such that $O \rrho C \ini A$ and $(\lambda
e.\ \lambda z.\ \pl{ind}(M)\ e)\ s\ O \rrho D = C$. Taking $C \equiv D$, the
first part follows trivially. Since $(\lambda e.\ \lambda z.\ \pl{ind}(M)\ e)\ s\ O \to^*
\pl{ind}(M)\ s \to M\ s\ (\lambda e.\ \lambda z.\ \pl{ind}(M)\ s)$, we get
the claim by Lemma \ref{realredclosed} and the inductive hypothesis.
\end{proof}

\begin{lemma}\label{realeqsymm}
There is a term $\pl{eqSymm}$ such that $\pl{eqSymm} \rrho \forall a, b.\ a
= b \to b = a$. 
\end{lemma}
\begin{proof}
Take
\[
\pl{eqSymm} \equiv \lambda a, b.\ \lambda x.\ N, \mbox{ where }
N=\pl{eqRep}(\lambda d.\
<\SND(\pl{eqProp}(x)\ d), \FST(\pl{eqProp}(x)\ d)>).
\]
To show that $\pl{eqSymm} \rrho \forall a, b.\ a = b \to b = a$, it suffices to show that for any 
$A, B, t, u, M$, if $M \rrho A = B$ then $N[x:=M] \rrho B = A$. Take any $A,
B, t, u, M$. The claim follows if for all $s \in T, C$ we can show:
\begin{enumerate}[$\bullet$]
\item There is $M_1$ such that $\SND(\pl{eqProp}(M)\ s) \downarrow \lambda
x.\ M_1$ and for all $N_1 \rrho C \ini B$, $M_1[x:=N_1] \rrho C \in A$.
\item There is $M_2$ such that $\FST(\pl{eqProp}(M)\ s) \downarrow \lambda
x.\ M_2$ and for all $N_2 \rrho C \ini A$, $M_2[x:=N_2] \rrho C \in B$. 
\end{enumerate}
Since $M \rrho A = B$, then there is $O$ such that $M \downarrow \pl{eqRep}(O)$, so $\FST(\pl{eqProp}(M)\ s) \to^* \FST(O\ s)$.
Moreover, for some $O_1, O_2$ we have $O\ s \downarrow <O_1, O_2>$, where
$O_1 \rrho C \ini A \to C \in B$ and $O_2 \rrho C \ini B \to C \in A$.
Therefore, $\FST(\pl{eqProp}(M)\ s)  \to^* O_1$ and similarly $\SND(\pl{eqProp}(M)\ s)  \to^* O_2$. We also know that
there are some $P_1, P_2$ such that $O_1 \downarrow \lambda x.\ P_1$, $O_2
\downarrow \lambda x.\ P_2$,  $P_1[x:=N_2] \rrho C \in B$ and $P_2[x:=N_1] \rrho C \in A$. Taking $M_1 =
P_2$ and $M_2 = P_1$, we get the claim by Lemma \ref{realredclosed}. 
\end{proof}

\begin{lemma}\label{realeqtrans}
There is a term $\pl{eqTrans}$ such that $\pl{eqTrans} \rrho \forall b, a,
c.\ a = b \land b = c \to a = c$. 
\end{lemma}
\begin{proof}
The proof and the realizers mirror closely the proof of Lemma \ref{eqtrans}. Set:
\begin{eqnarray*}
\pl{eqTrans} & = & \IND(M_0)\\
M_0 & = & \lambda b, x_1, a_1, c, x_2.\ \pl{eqRep}(\lambda f.\ <N, O>)\\
N & = & \lambda x_3.\ \LET\ [a_2, x_4] :=
\pl{inProp}(\FST(\pl{eqProp}(\FST(x_2))\ f)\ x_3)\ \IN\ N_1\\
N_1 &  = & \LET\ [a_3, x_5] := \pl{inProp}(\FST(\pl{eqProp}(\SND(x_2))\
a_2)\
\FST(x_4))\ \IN\ N_2\\
N_2 & = & \pl{inRep}([a_3, <\FST(x_5),  x_1\ a_2\ \FST(x_4)\ f\ a_3\
<\SND(x_4), \SND(x_5)>>])\\
O & = & \lambda x_3.\ \LET\ [a_2, x_4] :=
\pl{inProp}(\SND(\pl{eqProp}(\SND(x_2))\ f)\ x_3)\ \IN\ O_1\\
O_1 & = & \LET\ [a_3, x_5] := \pl{inProp}(\SND(\pl{eqProp}(\FST(x_2))\ a_2)\
\FST(x_4))\ \IN\ O_2\\
O_2 & = & \pl{inRep}([a_3, <\FST(x_5),  x_1\ a_2\ \FST(x_4)\ f\ a_3\
<\SND(x_4), \SND(x_5)>>]).
\end{eqnarray*}
We will show that for all $B$, $\pl{eqTrans} \downarrow
\lambda b.\ R$ for some term $R$ such that for any term $t$, $R[b:=t] \rrho
\forall a, c.\ a = B \land B = c \to a = c$, which trivially implies the claim. We proceed by induction on $\lambda$-rank of $B$.

We have $\pl{eqTrans} \to \lambda e.\ M_0\ e\ M_1$, where $M_1 = \lambda g.\ \lambda x.\ \pl{eqTrans}\ g$. Thus it suffices to show that
for all $t_1$, $M_0\ t_1\ M_1 \rrho \forall a, c.\ a = B \land B = c \to a
= c$. Since $M_0\ t_1\ M_1 \downarrow \lambda a_1, c, x_2.\
\pl{eqRep}(\lambda f.\ <N, O>[x_1:=M_1])$, it suffices to show
that for all $A, C, M_2$ such that $M_2 \rrho A = B \land B = C$ we have $\pl{eqRep}(\lambda f.\
<N, O>[x_1, x_2:=M_1, M_2]) \rrho A = C$. By Lemma \ref{eqin}, it suffices
to show that for all $F, u$ we have $N[x_1, x_2, f := M_1, M_2, u] \rrho F \ini A \to F \in C$ and
$O[x_1, x_2, f :=M_1, M_2, u] \rrho F \ini C \to F \in A$. 

For the proof of the first claim, we have $N[x_1, x_2, f :=M_1,M_2, u] \downarrow \lambda x_3.\ {\ldots}$. Take
any $M_3 \rrho F \ini A$. We need to show that:
\begin{eqnarray*}
\LET\ [a_2, x_4]& :=&
\pl{inProp}(\FST(\pl{eqProp}(\FST(M_2))\ u)\ M_3)\\
& \IN & N_1[x_1, x_2, x_3, f:=M_1, M_2,M_3, u] \rrho F \in C.
\end{eqnarray*}
We have $\FST(M_2) \rrho A = B$, so $\pl{eqProp}(\FST(M_2)) \rrho \forall f.\ (f \ini A \to f \in B) \land (f
\ini B \to f \in A)$, so by Lemma \ref{realimpl}
$\FST(\pl{eqProp}(\FST(M_2)\ u))\ M_3 \rrho F \in B$. Therefore,
$\FST(\pl{eqProp}(\FST(M_2)\ u))\ M_3 \downarrow \pl{inRep}(P)$ and $P
\downarrow [t_2, M_4]$ for
some $P, A_2, t_2, M_4$ such that $M_4 \rrho A_2 \ini B \land F = A_2$. Thus our term $\LET\ [a_2, x_4] := {\ldots} $ reduces to\footnote{Since
$x_3$ does not occur in $N_1$ and $N_2$, we omit it from the substitution.}
$N_1[x_1, x_2, x_4, a_2, f := M_1, M_2, M_4, t_2, u]$. 

Since $\SND(M_2) \rrho B = C$, we similarly have
$\FST(\pl{eqProp}(\SND(M_2))\ t_2)\ \FST(M_4) \rrho A_2 \in C$, so
$\FST(\pl{eqProp}(\SND(M_2))\ t_2)\
\FST(M_4) \downarrow \pl{inRep}(Q)$ and for some $A_3$, $Q \downarrow [t_3, M_5]$, $M_5
\rrho A_3 \ini C \land A_2 = A_3$. Therefore
\[
N_1[{\ldots}] \downarrow \pl{inRep}([t_3, <\FST(M_5), M_1\ t_2\ \FST(M_4)\ u\
t_3\ <\SND(M_4), \SND(M_5)>>])
\]
and by Lemma \ref{realredclosed} it suffices
to show that
\[
\pl{inRep}([t_3, <\FST(M_5), M_1\ t_2\ \FST(M_4)\ u\ t_3\ <\SND(M_4),
\SND(M_5)>>]) \rrho F \in C
\]
For this purpose, we need to show that $\FST(M_5) \rrho A_3 \ini C$, which
is trivial, and that
\[
M_1\ t_2\ \FST(M_4)\ u\ t_3\ <\SND(M_4), \SND(M_5)> \rrho F = A_3.
\]
Since $M_1 = \lambda g.\ \lambda x.\ \pl{eqTrans}\ g$, $\SND(M_4) \rrho F = A_2$ and $\SND(M_5)
\rrho A_2 = A_3$, all we need to have is that
$\pl{eqTrans}\ t_2 \rrho \forall a, c.\ a = A_2 \land A_2 = c \to a = c$. 
Since $\FST(M_4) \rrho A_2 \ini B$, $\lrk(A_2) \lt \lrk(B)$ and we get the claim by the inductive hypothesis.

The proof of the second claim proceeds in a very similar fashion. The only thing which
differs $O$ and $O_1$ from $N$ and $N_1$ is the exchange of $\FST$ and
$\SND$ which corresponds to using the information that $\forall f.\ f \ini C
\to f \in B$ and $\forall f.\ f \ini B \to f \in A$ and proceeding from $C$
to $A$ in the second part of the proof of Lemma \ref{eqtrans}. 
\ignore
{
The second case proceeds in a very similar fashion. We have $O[x_1:=M_1]
\downarrow \lambda x_3.\ {\ldots}$. Take any $M_3 \rrho F \ini C$. We have
$\SND(M_2) \rrho B = C$, so $\pl{eqProp}(\SND(M_2)) \rrho \forall f.\ (f
\ini B \to f \in C) \land (f \ini C \to f \in B)$, so $\LET [a_2, x_4] :=
\pl{inProp}(\SND((\pl{eqProp}(\SND(M_2))\ f))\ M_3)\ \IN\ O_1 \downarrow
N_1[a_2 := t_2, x_2 := M_2, x_4:=M_4]$ for some $t_2, A_2, M_4 \rrho
A_2 \ini B \land F = A_2$.

Since $\FST(M_2) \rrho A = B$, we similarly have
$\SND(\pl{eqProp}(\FST(M_2))\ t_2)\ \FST(M_4) \rrho A_2 \in A$, so
$\SND((\pl{eqProp}(\FST(M_2)) t_2))\
\FST(M_4) \downarrow \pl{inRep}(Q)$ and $Q \downarrow [t_3, M_5]$, $M_5
\rrho A_3 \ini A \land A_2 = A_3$ for some $A_3$. Thus
$O_1 \downarrow O_2[a_3:=t_3, x_5:=M_5]$. Thus it suffices to
show that $\pl{inRep}([t_3, <\FST(M_5), M_1\ t_1\ \FST(M_4)\ f\
a_3\ \SND(M_4)\ \SND(M_5)>]) \rrho F \in A$. For this purpose, it suffices
to show that $\FST(M_5) \rrho A_3 \ini A$, which is trivial, and that
$M_1\ t_1\ \FST(M_4)\ f\ a_3\ \SND(M_4)\ \SND(M_5) \rrho F = A_3$. 
Since $\FST(M_4) \rrho A_2 \ini B$, $\SND(M_4) \rrho F = A_2$ and $\SND(M_5)
\rrho A_2 = A_3$, all we need to have is that
$\pl{eqTrans} \rrho \forall a, c.\ a = A_2 \land A_2 = c \to a = c$. 
Since $\FST(M_4) \rrho A_2 \ini B$, by Lemma \ref{ineqrank},
$\lrk(A_2) \lt \lrk(B)$ and we get the claim by the inductive hypothesis.
}
\end{proof}
\begin{lemma}\label{leireal}
There is a term \pl{lei} such that $\pl{lei} \rrho \forall a, b, c.\ a \in c
\land a = b \to b \in c$. 
\end{lemma}
\begin{proof}
Take
\begin{eqnarray*}
\pl{lei} & = &\lambda a, b, c, x.\ \LET\ [d, y]:=\pl{inProp(\FST(x))}\
\IN\\
& & \pl{inRep}([d, <\FST(y), \pl{eqTrans}\ a\ b\ c\ <\pl{eqSymm}\ a\ b\
\SND(x), \SND(y)>>]).
\end{eqnarray*}
We need to show that for any $t_1, t_2, t_3 \in T$, $A, B, C$, for any $M
\rrho A \in C \land A = B$, we have
\begin{eqnarray*}
\LET\ [d, y]& :=& \pl{inProp(\FST(M))}\ \IN\\
& & \pl{inRep}([d, <\FST(y), \pl{eqTrans}\ t_1\ t_2\ t_3\ <\pl{eqSymm}\ t_1\ t_2\
\SND(M), \SND(y)>>]) \rrho B \in C.
\end{eqnarray*}
We have $M \downarrow <M_1, M_2>$, $M_1 \rrho A \in C$, $M_2 \rrho A = B$.
Therefore $M_1 \downarrow \pl{inRep(N)}$, $N \downarrow [u, O]$, $O \downarrow
<O_1, O_2>$ and there is $D$ such that $O_1 \rrho D \ini C$, $O_2 \rrho A =
D$. Therefore $\pl{inProp}(\FST(M)) \downarrow [u, O]$, so it suffices to
show that
\[
\pl{inRep}([u, <\FST(O), \pl{eqTrans}\ t_1\ t_2\ t_3\ <\pl{eqSymm}\ t_1\
t_2\ \SND(M), (\SND(O)>>]) \rrho B \in C.
\]
This follows if we can find some $E$ such that $O_1 \rrho E \ini
C$ and 
\[
\pl{eqTrans}\ t_1\ t_2\ t_3\ <\pl{eqSymm}\ t_1\ t_2\ \SND(M), \SND(O)>
\rrho B = E.
\] Take $E$ to be $D$. Since we have $\pl{eqSymm}\ t_1\ t_2\ \SND(M) \rrho
B = A$ and $\SND(O) \rrho A = E$, the claim follows by Lemma \ref{realeqtrans}.
\end{proof}

The following two lemmas will be used for the treatment of $\omega$ in Lemma
\ref{realsterms}.

\begin{lemma}\label{realunorderedpair}
If $A, B \in \vla$, then $\sr{\{ A, B \}} \in \vl_{\alpha + 1}$.
\end{lemma}
\begin{proof}
Take any $(M, C) \in \sr{\{ A, B \}}$. By the definition of $\sr{\{ A, B
\}}$, any such $C$ is in $\vla$, so $\sr{ \{ A, B \}} \in \vl_{\alpha + 1}$. 
\end{proof}

\begin{lemma}\label{realsucc}
If $A \in \vla$ and $M \rrho B = S(A)$, then $B \in \vl_{\alpha + 3}$.
\end{lemma}
\begin{proof}
$M \rrho B = S(A)$ means $M \rrho B = \bigcup \{ A, \{ A , A \} \}$.
By Lemma \ref{ineqrank}, it suffices to show that $\sr{\bigcup \{ A, \{ A ,
A \} \}} \in \vl_{\alpha + 3}$. Applying Lemma \ref{realunorderedpair}
twice, we find that $\sr{ \{ A, \{ A , A \} \}} \in \vl_{\alpha + 2}$. By
the definition of $\sr{\bigcup \{ A, \{ A , A \} \}}$, if $(M, C) \in
\sr{\bigcup \{ A, \{ A , A \} \}}$, then $C \in V_{\lrk(\sr{\bigcup \{ A, \{ A , A
\} \}})}$, so $C \in \vl_{\alpha + 2}$. Therefore $\sr{\bigcup \{ A, \{ A ,
A \} \}} \in \vl_{\alpha + 3}$ which shows the claim. 
\end{proof}

\begin{lemma}\label{realorderedpair}
If $A, B \in \vla$ and $M \rrho C = (A, B)$, then $C \in \vl_{\alpha + 2}$.
\end{lemma}
\begin{proof}
Similar to the proof of Lemma \ref{realsucc}, utilizing Lemmas \ref{realunorderedpair} and \ref{ineqrank}. 
\end{proof}

\begin{lemma}\label{lambdarank}
$\lrk(C) \leq rk(\rin{C}) + \omega$. 
\end{lemma}
\begin{proof}
If $(M, A) \in C$, then $M \rrho A \ini C$. We have $\pl{inRep}([a, <M,
\pl{eqRefl}\ a>]) \rrho A \in C$, so
$(\pl{inRep}([a, <M, \pl{eqRefl}\ a>]),  A) \in \rin{C}$. The extra $\omega$ is there to deal with possible difficulties with
finite $C$'s, as we do not know a priori the rank of set-theoretic encoding
of $\pl{inRep}([a, <M, \pl{eqRefl}\ a>]$. 
\end{proof}

\begin{lemma}\label{lll}
If $N \rrho \forall x \ino A.\ \phi$ then for all $(O, X) \in \rin{A}$, $N
\downarrow \lambda a.\ N_1$ and $N_1 \downarrow \lambda x.\ N_2$ and 
$N_2[x:=O] \rrho \phi[x:=X]$. Also, if $N \rrho
\exists x \ino A.\ \phi$ then there is $(O, X) \in \rin{A}$ such that $N
\downarrow [t, N_1]$, $N_1 \downarrow <O, N_2>$ and $N_2 \rrho \phi[x:=X]$.
\end{lemma}
\begin{proof}
If $N \rrho \forall x \ino A.\ \phi$ then $N \downarrow \lambda a.\ N_1$ and
for all $t, X$, $N_1[a:=t] \rrho X \in A \to \phi$. In particular, taking $t =
a$, we get $N_1 \downarrow \lambda x.\ N_2$ and
for all $O$ such that $O \rrho X \in A$, $N_2[x:=O] \rrho \phi[x:=X]$. This
implies that for all $X$, for all $O$, if $O \rrho X \ino A$, then $N \downarrow
\lambda a.\ N_1$, $N_1 \downarrow \lambda x.\ N_2$ and $N_2[x:=O] \rrho
\phi[x:=X]$, which proves the first part of the claim. 

If $N \rrho \exists x \ino A.\ \phi$, then $N \downarrow [t, N_1]$ and
there is $X$ such that $N_1 \downarrow <O, N_2>$, $O \rrho X \in A$ and
$N_2 \rrho \phi[x:=X]$, so there is $(O, X) \in \rin{A}$ such that $N
\downarrow [t, N_1]$, $N_1 \downarrow <O, N_2>$ and $N_2 \rrho \phi[x:=X]$. 
\end{proof}

With our lemmas in hand, we can now prove:

\begin{lemma}\label{vfunclosed}
Suppose $A \in \vis{i}$ and $N \rrho $''$C$ is a function from $A$ into $V_i$''. Then
$C \in \vinac{i}$.
\end{lemma}
\begin{proof}
First let us write formally the statement ``$C$ is a function from $A$ into
$V_i$''. This means ``for all $x \in A$ there is exactly one $y \in V_i$
such that $(x, y) \in C$ and for all $z \in C$ there is $x \in A$ and $y \in
V_i$ such that $z = (x, y)$''. Thus $N \downarrow <N_1, N_2>$, $N_1 \rrho 
\forall x \ino A \exists !y \ino V_i.\ (x, y) \ino C$ and $N_2 \rrho \forall z
\ino C \exists x \ino A \exists y \ino V_i.\ z \eqo (x, y)$. So $N_1 \rrho
\forall x \ino A \exists y \ino V_i.\ (x, y) \ino C \land \forall z.\ (x, z)
\in C \to z = y$. By Lemma \ref{lll}, for all $(O, X) \in \rin{A}$ there is $(P, Y) \in
\rin{\vis{i}}$ such that $\phi(O, X, P, Y)$ holds, where $\phi(O, X, P, Y)$ is defined as:
\begin{eqnarray*}
\phi(O, X, P, Y) & \equiv & (N_1 \downarrow \lambda a.\ N_{11}) \land
(N_{11} \downarrow \lambda x.\ N_{12}) \land (N_{12}[x:=O] \downarrow [t, N_{13}]) \land \\
& & (N_{13} \downarrow <P, Q>) \land (Q \downarrow <Q_1, Q_2>) \land \\
& & (Q_1 \rrho (X, Y) \ino C) \land (Q_2 \rrho \forall z.\ (X, z) \ino C \to z
\eqo Y)
\end{eqnarray*}
Let $\psi(O, X, P, Y)$ be defined as:
\[
\psi(O, X, P, Y) \equiv \exists Q_1, Q_2.\ (Q_1 \rrho (X, Y) \ino C) \land
(Q_2 \rrho \forall z.\ (X, z) \ino C \to z = Y)
\]
Obviously, if $\phi(O, X, P, Y)$ then $\psi(O, X, P, Y)$. So for all $(O, X)
\in \rin{A}$ there is $(P, Y) \in \rin{\vis{i}}$ such that $\psi(O, X, P,
Y)$ holds. 

Define a function $F$ which takes $(O, X) \in \rin{A}$ and returns $\{ (P,
Y) \in \rin{\vis{i}}\ |\ \psi(O, X, P, Y) \}$. Suppose $(P_1, Y_1), (P_2,
Y_2) \in F((O, X))$. Then there are $Q_{11}, Q_{12}, Q_{21}$ such that
$Q_{11} \rrho (X, Y_1) \in C$, $Q_{12} \rrho \forall z.\ (X, z) \in C \to z
= Y_1$, $Q_{21} \rrho (X, Y_2) \in C$. By Lemma
\ref{lll}, $Q_{12} \downarrow \lambda a.\ R_1$, $R_1 \downarrow \lambda x.\
R_2$ and $R_2[x:=Q_{21}] \rrho Y_2 = Y_1$. Since $\pl{eqSymm}\ a\ a\ R_2[x:=Q_{21}] \rrho Y_1 = Y_2$, by Lemma \ref{ineqrank} the $\lambda$-ranks of $Y_1, Y_2$ are
the same and, since any such $(P, Y)$ is a member of $\rin{\vis{i}}$, they are
smaller than \inac{i}. Also, for any $(O, X) \in \rin{A}$, $F(O, X)$ is
inhabited. 

Furthermore, define a function $G$ from $\rin{A}$ to $\inac{i}$, which takes $(O, X)
\in \rin{A}$ and returns $\bigcup \{ \lrk((P, Y))\ |\ (P, Y) \in F(O, X) \land
\psi(O, X, P, Y) \}$. Then for any $(O, X) \in \rin{A}$, $G(O, X)$ is an
ordinal smaller than $\inac{i}$ and if $(P, Y) \in \rin{\vis{i}}$ and
$\psi(O, X, P, Y)$, then $(P, Y) \in V^\lambda_{G(O, X)}$. Moreover, as
$\inac{i}$ is inaccessible, $G \in R(\inac{i})$, where $R(\inac{i})$ denotes
the $\inac{i}$-th element of the standard cumulative hierarchy. Therefore $\bigcup ran(G)$ is also an
ordinal smaller than $\inac{i}$. We define an ordinal $\beta$ to be
$\max(\lrk(A), \bigcup ran(G))$. 

Now take any $(M, B) \in \rin{C}$, so $M \rrho B \in C$. Then, by the
definition of $N_2$ and Lemma \ref{lll} there is $(O, X) \in \rin{A}$ and $(O_1,
Z) \in \rin{\vis{i}}$ such that $N_2 \downarrow \lambda a.\ N_{21}$, $N_{21}
\downarrow \lambda x.\ N_{22}$, $N_{22}[x:=M] \downarrow [t, N_{23}]$,
$N_{23} \downarrow <O, N_{24}>$, $N_{24} \downarrow [t, N_{25}]$, $N_{25}
\downarrow <O_1, R>$ and $R \rrho B = (X, Z)$. Let $M_1 = \pl{lei}
\ a\ a\ a\ <M, R>$, then $M_1 \rrho (X, Z) \in C$. Take any element $(P, Y) \in F(O, X)$
and accompanying $Q_1, Q_2$. Then $Q_2 \downarrow \lambda a.\ Q_3$, $Q_3
\downarrow \lambda x.\ Q_4$ and $Q_4[x:=M_1] \rrho Z = Y$. By Lemma \ref{ineqrank}, $\lrk(Z) \leq \lrk(Y)$ and
thus $\lrk(Z) \leq \beta$. Since $(O, X) \in \rin{A}$, $\lrk(X) \leq \beta$, too.
By Lemma \ref{realorderedpair}, $\lrk(B) \leq \beta + 2$. By Lemma
\ref{lambdarank}, $rk(B) \leq \beta + \omega$, so $rk(\rin{C}) \leq \beta +
\omega + 1$. By Lemma \ref{lambdarank} again, $\lrk(C) \leq \beta + 2\omega$.
Since $\beta+2\omega$ is still smaller than $\inac{i}$, we get the claim. 
\end{proof}

\begin{lemma}\label{visin}
If $M \rrho A \in \vis{i, \gamma}$, then $M \rrho A \in V_i$.
\end{lemma}
\begin{proof}
If $M \rrho A \in \vis{i, \gamma}$, then $M \downarrow \pl{inRep}(N)$, $N
\downarrow [t, O]$, $O \downarrow <O_1, O_2>$ and there is
$C$ such that $O_1 \downarrow v$, $(v, C) \in
\vis{i, \gamma}$, $O_2 \rrho C = A$. Then also $(v, C) \in \vis{i}$, so $O_1
\rrho C \ini V_i$, so also $M \rrho A \in V_i$. 
\end{proof}

\begin{lemma}\label{visclauses}
If $N \rrho \psi_i(C, \vis{i, \gamma})$, where $\psi_i$ is one of the five
clauses defining $\phiinaca(C, \vis{i, \gamma})$ in the Definition
\ref{dinac}, then $N \rrho \psi_i(C, V_i)$.
\end{lemma}
\proof
There are five cases to consider:
\begin{enumerate}[$\bullet$]
\item $N \rrho C = V_{i-1}$. This case is trivial.
\item $N \rrho \exists a.\ a \in \vis{i, \gamma} \land c \in a$. Then there is $A$ such that $N
\downarrow [t, O]$, $O \downarrow <O_1, O_2>$, $O_1 \rrho A \in \vis{i,
\gamma}$, $O_2 \rrho C \in A$. By Lemma \ref{visin}, $O_1 \rrho A \in V_i$,
so also $N \rrho \exists a.\ a \in V_i \land c \in a$. 
\item $N \rrho \exists a.\ a \in \vis{i, \gamma} \land c = \bigcup a$. 
Then there is $A$ such that $N \downarrow [t, O]$, $O \downarrow <O_1,
O_2>$, $O_1 \rrho A \in
\vis{i, \gamma}$, $O_2 \rrho C = \bigcup A$. Thus by Lemma \ref{visin} $O_1
\rrho A \in V_i$ and we get the claim in the same way as in the previous
case. 
\item $N \rrho \exists a.\ a \in \vis{i, \gamma} \land C = P(a)$. Similar to the previous case.
\item $N \rrho \exists a.\ a \in \vis{i, \gamma} \land C \in a \to \vis{i,
\gamma}$. Then there is $A$ such that $N \downarrow [t, O]$, $O \downarrow
<O_1, O_2>$, $O_1 \rrho
A \in \vis{i, \gamma}$, $O_2 \rrho ``$$C$ is a function from $A$ into
\vis{i, \gamma}''. By Lemma \ref{visin}, $O_1 \rrho A \in V_i$. Expanding
the second part, we have $O_2 \downarrow <P_1, P_2>$,
$P_1 \rrho  \forall x \ino A \exists !y \ino \vis{i, \gamma}.\ (x, y) \ino
C$ and $P_2 \rrho \forall z \ino C \exists x \ino A \exists y \ino \vis{i, \gamma}.\ z \eqo (x, y)$.
We will tackle $P_1$ and $P_2$ separately.
\begin{enumerate}[-]
\item For $P_1$, we have for all $X, t$, $P_1 \downarrow \lambda a.\ P_{11}$,
$P_{11}[a:=t] \downarrow \lambda x. Q$ and for all $R \rrho X \in A$ there is $Y$
such that $Q[x:=R] \downarrow [t_1, Q_0]$, $Q_0 \downarrow <Q_1, Q_2>$, $Q_1 \rrho Y \in \vis{i, \gamma}$
and $Q_2 \rrho (X, Y) \in C \land \forall z.\ (X, z) \in C \to z = Y$. By
Lemma \ref{visin} we also have $Q_1 \rrho Y \in V_i$, so also $P_1 \rrho
\forall x \in a \exists! y.\ y \in V_i \land (x, y) \in C$. 
\item For $P_2$, we have for all $Z, t$, $P_2 \downarrow \lambda a.\ P_{11}$,
$P_{11}[a:=t] \downarrow \lambda x. Q$ and for all $R
\rrho Z \in C$ there are $X, Y$ such that $Q[x:=R] \downarrow [t_1, Q_0]$,
$Q_0 \downarrow <Q_1, Q_2>$ and $Q_1 \rrho X \in A$. Moreover, 
$Q_2 \downarrow [t_2, S_0]$, $S_0 \downarrow
<S_1, S_2>$ and $S_1 \rrho Y \in \vis{i, \gamma}$. By Lemma \ref{visin}
we also have $S_1 \rrho Y \in V_i$, so also $P_2 \rrho \forall z \in C \to \exists x \in A \exists y \in V_i.\ z = (x, y)$. 
\end{enumerate}
Therefore also $O_2 \rrho$ ``$C$ is a function from $A$ into $V_i$'' and in
the end $N \rrho \exists a.\ a \in V_i \land C \in a \to V_i$.\qed
\end{enumerate}

\begin{corollary}\label{visinaca}
If $M \rrho \phiinaca(C, \vis{i, \gamma})$, then $M \rrho \phiinaca(C, V_i)$.
\end{corollary}

The following lemma states the crucial property of the realizability relation.

\begin{lemma}\label{realsterms}
$(M, C) \in \SB{t_A(\ov{u})}_\rho$ iff $M = \pl{axRep}(N)$ and $N
\reals_\rho \phi_A(C, \overrightarrow{\SB{u}_\rho)}$.
\end{lemma}

\begin{proof}
The proof proceeds by case analysis on $t_A(\ov{u})$. We first do the proof
for all terms apart from $\omega$ and $V_i$, then we show the claim for
$\omega$ and finally for $V_i$.

For all terms, save
$\omega$ and $V_i$, the left-to-right direction is immediate. For the right-to-left direction,
suppose $N \reals_\rho \phi_A(C, \overrightarrow{\SB{u}_\rho})$ and $M = \pl{axRep}(N)$. To
show that $(M, C) \in \SB{t_A(\ov{u})}_\rho$, we need to show that $C
\in \vl_\gamma$. Let $\ov{\alpha} = \overrightarrow{rank(\SB{u}_\rho)}$. Case $t_A(\ov{u})$ of:
\begin{enumerate}[$\bullet$]
\item $\{ u_1, u_2 \}$. Suppose that $N \reals_\rho C = \SB{u_1}_\rho \lor C
= \SB{u_2}_\rho$. Then either $N \downarrow\ \INL(N_1) \land N_1 \reals_\rho C =
\SB{u_1}_\rho$ or $N \downarrow\ \INR(N_1) \land N_1 \reals_\rho C =
\SB{u_2}_\rho$. By Lemma \ref{ineqrank}, in the former case $C \in
\vl_{\alpha_1}$, in the latter $C \in \vl_{\alpha_2}$, so $C \in
\vl_{max(\alpha_1, \alpha_2)}$. 
\item $P(u)$. Suppose that $N \reals_\rho \forall d.\ d\ \in C \to d \in
\SB{u}_\rho$. Then $N \downarrow \lambda a.\ N_1$ and for any $t$, $\forall
D.\ N_1[a:=t] \reals_\rho D \in C \to D \in \SB{u}_\rho$, so $\forall D, t.\
N_1[a:=t] \downarrow \lambda x.\ N_2$ and for all $O$, if $O \reals D \in C$
then $N_2[x:=O] \reals_\rho D \in \SB{u}_\rho$. Take any $(v, B) \in C$.
Then $\pl{inRep}([a, <v, \pl{eqRefl\ a}>]) \rrho B \in C$, so
$N_2[x:=\pl{inRep}([a, <v, \pl{eqRefl\ a}>]] \rrho B \in \SB{u}_\rho$.
Thus by Lemma \ref{ineqrank} any such $B$ is in $\vl_{\alpha}$, so $C \in \vl_{\alpha + 1}$.
\item $\bigcup u$. Suppose $N \reals_\rho \exists c.\ c \in \SB{u}_\rho
\land C \in c$. Then $N \downarrow [t, N_1]$ and there is $B$ such that $N_1
\rrho B \in \sr{u} \land C \in B$. Thus $N_1 \downarrow <N_1, N_2>$, $N_1
\rrho B \in \sr{u}$, $N_2 \rrho C \in B$. By Lemma \ref{ineqrank}, any such
$B$ is in $\vl_{\alpha}$, so also $C \in \vl_{\alpha}$. 
\item $S_{\phi(a, \ov{f})}(u, \ov{u})$.  Suppose $N \reals_\rho C \in
\SB{u}_\rho \land \phi(C, \overrightarrow{\sr{u}})$. Then $N \downarrow <N_1, N_2>$ and
$N_1 \rrho C \in \sr{u}$. Thus $C \in \vl_{\alpha_1}$.
\item $R_{\phi(a, \ov{f})}(u, \ov{u})$. Suppose $N \reals_\rho (\forall x \in \SB{u}_\rho \exists! y.\ \phi(x,
y, \overrightarrow{\SB{u}_\rho})) \land \exists x \in \SB{u}_\rho.\ \phi(x,
C, \overrightarrow{\SB{u}_\rho})$. Then $N \downarrow
<N_1, N_2>$ and $N_2 \reals_\rho\exists x \in \SB{u}_\rho.\ \phi(x, C,
\overrightarrow{\SB{u}_\rho})$. Thus $N_2 \downarrow [t, N_{20}]$, $N_{20} \downarrow
<N_{21}, N_{22}>$ and there is $B$ such that $N_{21} \reals_\rho B \in
\SB{u}_\rho$ and $N_{22} \reals_\rho\phi(B, C, \overrightarrow{\SB{u}_\rho})$. We also
have $N_1 \reals_\rho\forall x \in \SB{u}_\rho \exists! y.\ \phi(x, y,
\overrightarrow{\SB{u}_\rho})$, so $N_1 \downarrow \lambda a.\ N_{11}$ and for all $C$, $N_{11} \downarrow
\lambda x.\ O$ and for all $P \rrho C \in \SB{u}_\rho$, $O[x:=P]
\rrho \exists !y.\ \phi(C, y, \overrightarrow{\SB{u}_\rho})$. So taking $C = B$
and $P=N_{21}$, there is $D$ such that $N_1 \downarrow \lambda a.\ N_{11}$,
$N_{11} \downarrow \lambda x.\ O$ and $O[x:=N_{21}] \downarrow [s, O_1]$ and
$O_1 \rrho \phi(B, D, \overrightarrow{\SB{u}_\rho}) \land \forall e.\
\phi(B, e, \overrightarrow{\SB{u}_\rho}) \to e =
D$. Therefore $(N_1, (N_{21}, B)) \in G$ from the definition of $\gamma$, so 
there is $D \in V^{\lambda}_\gamma$ such that $N_1 \downarrow \lambda a.\
N_{11}$, $N_{11} \downarrow \lambda x.O$, $O[x:=N_{21}] \downarrow [s, O_1]$ and $O_1 \reals_\rho \phi(B,
D, \overrightarrow{\SB{u}_\rho}) \land \forall e.\ \phi(B, e, \overrightarrow{\SB{u}_\rho}) \to e =
D$. So $O_1 \downarrow <O_{11}, O_{12}>$ and $O_{12} \reals_\rho\forall e.\
\phi(B, e, \overrightarrow{\SB{u}_\rho}) \to e = D$. Therefore, $O_{12} \downarrow
\lambda a.\ Q$, $Q \downarrow \lambda x.\ Q_1$ and $Q_1[x:=N_{22}]
\reals_\rho C = D$. By Lemma \ref{ineqrank}, $C \in \vl_\gamma$.
\end{enumerate}
Now we tackle $\omega$. For the left-to-right direction, obviously $M =
\pl{infRep}(N)$. For the claim about $N$ we proceed by induction on the
definition of $\omega'$:
\begin{enumerate}[$\bullet$]
\item The base case. Then $N \downarrow \INL(O)$ and $O \reals_\rho A = 0$, so $N
\reals_\rho A = 0 \lor \exists y \in \omega'.\ A = S(y)$. 
\item Inductive step. Then $N \downarrow \INR(N_1)$, $N_1 \downarrow [t, O]$,
$O \downarrow <M', P>$, $(M', B) \in \omega'^+$, $P \reals_\rho A = S(B)$.
Therefore, there is $C$ (namely $B$) such that $M' \reals_\rho C \in \omega'$ and $P
\reals_\rho A = S(C)$. Thus $[t, O] \reals_\rho \exists y.\ y \in
\omega' \land A = S(y)$, so $N \reals_\rho A = 0 \lor \exists y \in \omega'.\ A = S(y)$. 
\end{enumerate}
For the right-to-left direction, suppose $N \reals_\rho A = 0 \lor (\exists y.\ y \in
\omega'\land A = S(y))$. Then either $N \downarrow
\INL(N_1)$ or $N \downarrow \INR(N_1)$. In the former case, $N_1 \reals_\rho A =
0$, so by Lemma \ref{ineqrank} $A \in \vl_\omega$. In the latter, $N_1 \reals_\rho\exists y.\ y \in
\omega' \land A = S(y)$. Thus $N_1 \downarrow [t, O]$ and there is $B$ such that $O
\reals_\rho B \in \omega' \land A = S(B)$. So $O
\downarrow <M', P>$, $(M', B) \in \omega'^+$ and $P \reals_\rho A =
S(B)$. This is exactly the inductive step of the
definition of $\omega'$, so it remains to show that $A \in
\vl_\omega$. Since $(M', B) \in \omega'^+$, there is a finite ordinal
$\alpha$ such that $B \in \vl_\alpha$. By Lemma \ref{realsucc}, $A \in
\vl_{\alpha + 3}$, so also $A \in \vl_\omega$ and we get the claim. 

Finally, we take care of $V_i$. We first show the left-to-right direction. Suppose $(M, A) \in
\vis{i}$, then $M = \pl{inac_iRep}(N)$. We must have $N \rrho \phiinaca(A,
\vis{i, \gamma}) \land \forall d.\ \phiinacb(d) \to A \in d$ for some ordinal
$\gamma$. Then $N \downarrow <N_1, N_2>$, $N_1 \rrho \phiinaca(A,
\vis{i, \gamma})$, $N_2 \rrho \forall d.\ \phiinacb(d) \to A \in d$. Corollary
\ref{visinaca} gives us $N_1 \rrho \phiinaca(A, V_i)$, so $N \rrho \phiinaca(A,
V_i) \land \forall d.\ \phiinacb(d) \to A \in d$, which is what we want. 

For the right-to-left direction, suppose $N \rrho \phiinaca(C,
V_i) \land \forall d.\ \phiinacb(d) \to C \in d$. We need to show that
$(\pl{inac_iRep(N)}, C) \in \vis{i}$. By the definition of \vis{i} it suffices to
show that $C \in V_{\inac{i}}$. We have $N \downarrow <N_1, N_2>$ 
and $N_1 \rrho $ ``$C$ is equal to $V_{i-1}$ or there is $A \in V_i$ such that
$C$ is a powerset/union/member of $A$, or $C$ is a function from $A$ into
$V_i$.''. The proof splits into corresponding five cases. The first four are easy to prove using Lemma
\ref{ineqrank} and the definition of the ordinal $\gamma$ in the clause
\ref{termdef} in the definition of realizability. The last one follows by 
Lemma \ref{vfunclosed}.
\end{proof}

\section{Normalization}\label{sectionnorm}

In this section, environments $\rho$ are finite partial functions mapping 
propositional variables to terms of $\la$ and first-order variables to pairs $(t,
A)$, where $t \in T$ and $A \in \vl$. Therefore, $\rho : Var \cup FVar \to
\lat \cup (T \times \vl)$, where $Var$ denotes the set of propositional variables
and $FVar$ denotes the set of first-order variables. Note that any $\rho$ can be used as a realizability environment by considering
only the mapping of first-order variables to $\vl$. Therefore we will be
using the notation $\rrho$ also for these environments $\rho$.

\begin{definition}
For a sequent $\gp M : \phi$, $\rho \models \gp M : \phi$ means that $\rho$ is
defined on $FV(\Gamma, M, \phi)$ and for all $(x_i, \phi_i) \in \g$, $\rho(x_i) \reals_\rho \phi_i$.
\end{definition}

Note that if $\rho \models \gp M : \phi$, then for any term $t$ in $\g, \phi$,
$\SB{t}_{\rho}$ is defined and so is the realizability relation $M
\reals_{\rho} \phi$.

\begin{definition}
For a sequent $\gp M : \phi$, if $\rho \models \gp M : \phi$ then $M[\rho]$
is $M[x_1 := \rho(x_1), {\ldots} , x_n := \rho(x_n), a_1:=\rho_T(a_1),
{\ldots}, a_k:=\rho_T(a_k)]$, where $FV(M) = \{ x_1, {\ldots}, x_n \}$,
$FV_F(M) = \{ a_1, {\ldots} , a_k \}$ and $\rho_T$ denotes the restriction of $\rho$ to the mapping from first-order
variables into terms: $\rho_T = \lambda a \in FVar.\ \pi_1(\rho(a))$.
\end{definition}

\begin{lemma}\label{rhosubst}
$M[\rho][x:=N] = M[\rho[x:=N]]$. Also $M[\rho][a:=t] = M[\rho[a:=(t, A)]]$. 
\end{lemma}
\begin{proof}
Straightforward structural induction on $M$. 
\end{proof}

\begin{thm}[Normalization]\label{norm}
If $\gp M : \vartheta$ then for all $\rho \models \gp M : \vartheta$, $\overline{M}[\rho] \reals_\rho \vartheta$.
\end{thm}
\proof 
For any $\lii$ term $M$, $M'$ in the proof denotes $\overline{M}[\rho]$.
We proceed by metalevel induction on $\gp M : \vartheta$. Case $\gp M : \vartheta$ of:
\begin{enumerate}[$\bullet$]
\item 
\[
\infer{\g, x : \phi \proves x : \phi}{}
\]
Then $M' = \rho(x)$ and the claim follows.
\item 
\[
\infer{\gp M\ N : \psi}{\gp M : \phi \to \psi & \gp N : \phi}
\]
By the inductive hypothesis, $M' \reals_\rho \phi \to \psi$ and $N' \reals_\rho \phi$. Lemma
\ref{realimpl} gives the claim.
\item
\[
\infer{\gp \lambda x : \phi.\ M : \phi \to \psi}{\g, x : \phi \p M : \psi}
\]
We need to show that for any $N \reals_\rho \phi$, $M'[x:=N] \reals_\rho \psi$. Take any such $N$. Let 
$\rho' = \rho[x:=N]$. Then $\rho' \models \Gamma, x : \phi \p M : \psi$, so by the
inductive hypothesis $\overline{M}[\rho'] \reals_{\rho'} \psi$. By Lemma
\ref{rhosubst}  $\overline{M}[\rho'] = \overline{M}[\rho][x:=N] = M'[x:=N]$, so $M'[x:=N] \reals_{\rho'}
\psi$. Since $\rho'$ agrees with $\rho$ on logic variables, by Lemma \ref{afvreal} we get $M'[x:=N] \reals_\rho \psi$.
\item 
\[
\infer{\gp \pl{magic}(M) : \phi}{\gp M : \bot}
\]
By the inductive hypothesis, $M' \rrho \bot$, which is not the case, so
anything holds, in particular $\pl{magic}(M') \reals_\rho \phi$.
\item
\[
\infer{\gp \FST(M) : \phi}{\gp M : \phi \land \psi}
\]
By the inductive hypothesis, $M' \reals_\rho \phi \land \psi$, so $M' \downarrow <M_1, M_2>$ and
$M_1 \reals_\rho \phi$. Therefore $\FST(M) \to^* \FST(<M_1, M_2>) \to M_1$.
Lemma \ref{realredclosed} gives the claim. 
\item 
\[
\infer{\gp \SND(M) : \psi}{\gp M : \phi \land \psi}
\]
Symmetric to the previous case. 
\item 
\[
\infer{\gp <M, N> : \phi \land \psi}{\gp M : \phi & \gp N : \psi}
\]
All we need to show is $M' \reals_\rho \phi$ and $N' \reals_\rho \psi$, which we
get from the inductive hypothesis.
\item
\[
\infer{\gp \INL(M) : \phi \lor \psi}{\gp M : \phi}
\]
We need to show that $M' \reals_\rho \phi$, which we get from the inductive hypothesis.
\item
\[
\infer{\gp \INR(M) : \phi \lor \psi}{\gp M : \psi}
\]
Symmetric to the previous case.
\item 
\[
\infer{\gp \CASE(M, x : \phi.\ N, x : \psi.\ O) : \vartheta}{\gp M :
\phi \lor \psi & \g, x : \phi \proves N : \vartheta & \g, x : \psi \proves O : \vartheta}
\]
By the inductive hypothesis, $M' \rrho \phi \lor \psi$. Therefore either $M'
\downarrow \INL(M_1)$ and $M_1 \rrho \phi$ or $M' \downarrow \INR(M_2)$ and
$M_2 \rrho \psi$. We only treat the former case, the latter is symmetric.
Since $\rho[x:=M_1] \rrho \g, x : \phi \proves N : \vartheta$, by the
inductive hypothesis we get $\overline{N}[\rho[x:=M_1]] \rrho \vartheta$. We also
have $\CASE(M, x.\overline{N}, x.\overline{O}) \to^* \CASE(\INL(M_1), x.\overline{N}, x.\overline{O}) \to
\overline{N}[x:=M_1]$. By Lemma \ref{rhosubst}, $\overline{N}[x:=M_1] =
\overline{N}[\rho[x:=M_1]]$, so Lemma \ref{realredclosed} gives us the claim.
\item
\[
\infer{\gp \lambda a.\ M : \forall a.\ \phi}{\gp M : \phi}
\]
By the inductive hypothesis, for all $\rho \models \gp M : \phi$, $\overline{M}[\rho]
\reals \phi$. We need to show that for all $\rho \models \gp \lambda a.\ M :
\forall a.\ \phi$, $\overline{(\lambda a.\ M)}[\rho] \reals_\rho \forall a.\
\phi$. This is equivalent to $\lambda a.\ \overline{M}[\rho] \rrho \forall a.\
\phi$. Take any such $\rho$. We need to show that $\forall A, t.\ \overline{M}[\rho][a:=t] \reals_\rho
\phi[a:=A]$. Take any $A$ and $t$. Since $\rho[a:=(t, A)] \models \gp M :
\phi$ and by Lemma \ref{rhosubst} $\overline{M}[\rho][a:=t] = \overline{M}[\rho[a:=(t, A)]]$, we get the claim by the inductive hypothesis.
\item
\[
\infer{\gp M\ t : \phi[a:=t]}{\gp M : \forall a.\ \phi}
\]
By the inductive hypothesis, $M' \reals_\rho \forall a.\ \phi$, so $M' \downarrow \lambda a.\ N$
and $\forall A, u.\ N[a:=u] \reals_\rho \phi[a:=A]$. In particular $N[a:=t[\rho]]
\rrho \phi[a:=\sr{t}]$. By Lemma \ref{realsubst}, $N[a:=t[\rho]] \reals_\rho
\phi[a:=t]$. Since $M'\ (t[\rho]) \to^* (\lambda a.\ N)\ t[\rho] \to
N[a:=t[\rho]]$, Lemma \ref{realredclosed} gives us  the claim.
\item 
\[
\infer{\gp [t, M] : \exists a.\ \phi}{\gp M : \phi[a:=t]}
\]
By the inductive hypothesis, $M' \reals_\rho \phi[a:=t]$, so by Lemma \ref{realsubst},
$M' \reals_\rho \phi[a:=\SB{t}_\rho]$. Thus, there is a lambda-name $A$, namely $\SB{t}_\rho$, such that $M' \reals_\rho \phi[a:=A]$. Thus, 
$\overline{[t, M]}[\rho]=[t[\rho], M'] \reals_\rho \exists a.\ \phi$ which is what we want.
\item
\[
\infer[a \notin FV(\Gamma, \psi)]{\gp \LET\ [a, x : \phi] := M\ \IN\ N : \psi}
{\gp M : \exists a.\ \phi & \g, x : \phi \proves N : \psi}
\]
Let $\rho \models \gp \LET\ [a, x : \phi] := M\ \IN\ N : \psi$. We need to
show $\overline{\LET\ [a, x : \phi ] := M\ \IN\ N}[\rho] = \LET\ [a, x] := M'\ \IN\ \overline{N}[\rho] \rrho \psi$.
By the inductive hypothesis, $M' \rrho \exists a.\ \phi$, so $M' \downarrow [t, M_1]$ and
for some $A$, $M_1 \rrho \phi[a:=A]$. By the inductive hypothesis again, for any $\rho' \models \g,
x : \phi \p N : \psi$ we have $\overline{N}[\rho'] \reals_{\rho'} \psi$. Take
$\rho' = \rho[x:=M_1, a:=(t, A)]$. Since $a \notin FV(\psi)$, by Lemma
\ref{afvreal} $\overline{N}[\rho'] \rrho \psi $. Now, $\LET\ [a, x : \phi] := M'\
\IN\ \overline{N}[\rho] \to^* \LET\ [a, x] :=
[t, M_1]\ \IN\ \overline{N}[\rho] \to \overline{N}[\rho][a:=t][x:=M_1] = \overline{N}[\rho']$.
Lemma \ref{realredclosed} gives us the claim.
\item 
\[
\infer{\gp \pl{eqRep}(t, u, M) : t \eqo u}{\gp M : \forall d.\ (d \ini t \to d
\ino u) \land (d \ini u \to d \ino t)}
\]
By the inductive hypothesis, $M' \rrho \forall d.\ (d \ini t \to d \ino u) \land (d \ini u \to d
\ino t)$. By Lemma \ref{realsubst}, $M' 
\rrho \forall d.\ (d \ini \sr{t} \to d \ino \sr{u}) \land (d \ini \sr{u} \to d
\ino \sr{t})$. By Lemma \ref{eqin}, $\pl{eqRep}(M') \rrho \sr{t} \eqo \sr{u}$.
Lemma \ref{realsubst} applied again gives us the claim. 
\[
\infer{\gp \pl{eqProp}(t, u, M) : \forall d.\ (d \ini t \to d
\ino u) \land (d \ini u \to d \ino t)}{\gp M : t \eqo u}
\]
By the inductive hypothesis, $M' \rrho t \eqo u$. By Lemma \ref{realsubst}, $M' \rrho \sr{t} \eqo
\sr{u}$. By Lemma \ref{eqin}, $M' \downarrow \pl{eqRep}(N)$ and
$N \rrho \forall d.\ (d \ini \sr{t} \to d \ino \sr{u}) \land (d \ini \sr{u}
\to d \ino \sr{t})$. Since $\overline{\pl{eqProp}(t, u, M)} = \pl{eqProp}(M') \to^*
\pl{eqProp}(\pl{eqRep}(N)) \to N$, by Lemma \ref{realredclosed}
$\overline{\pl{eqProp}(t, u, M)} \rrho \forall d.\ (d \ini \sr{t} \to d \ino \sr{u}) \land (d \ini \sr{u}
\to d \ino \sr{t})$. Lemma \ref{realsubst} applied once again gives us the claim.
\item For $\pl{inProp}$ and $\pl{inRep}$, the proof is similar to the two
previous cases. 
\item 
\[
\infer{\gp \pl{axRep}(t, \ov{u}, M) : t \ini t_A(\ov{u})}{\gp M : \phi_A(t, \ov{u})}
\]
By the inductive hypothesis, $M' \reals_\rho \phi_A(t, \ov{u})$. By Lemma \ref{realsubst} 
this is equivalent to $M' \reals_\rho \phi_A(\SB{t}_\rho, \overrightarrow{\SB{u}_\rho})$.
By Lemma \ref{realsterms} $(\pl{axRep}(M'), \SB{t}_\rho) \in
\SB{t_A(\ov{u})}_\rho$, so $\pl{axRep}(M') \rrho
t \ini t_A(\ov{u})$. 
\item
\[
\infer{\gp \pl{axProp}(t, \ov{u}, M) : \phi_A(t, \ov{u}) }{ \gp M : t
\ini t_A(\ov{u})}
\]
By the inductive hypothesis, $M' \reals_\rho t \ini t_A(\ov{u})$. This means that $M' \downarrow v$ and
$(v, \SB{t}_\rho) \in \sr{t_A(\ov{u})}$. By Lemma \ref{realsterms}, $v =
\pl{axRep}(N)$ and $N \reals_\rho \phi_A(\SB{t}_\rho, \overrightarrow{\SB{u}_\rho})$.
By Lemma \ref{realsubst}, $N \reals_\rho \phi_A(t, \ov{u})$.
Moreover, $\overline{\pl{axProp}(t, \ov{u}, M)} = \pl{axProp}(M') \to^*
\pl{axProp}(\pl{axRep}(N)) \to
N$. Lemma \ref{realredclosed} gives us the claim.
\item
\[
\infer{\gp \IND(M, \ov{t}) : \forall a.\
\phi(a, \ov{t})}{\gp M : \forall c.\ (\forall b.\ b \ini c \to \phi(b,
\ov{t})) \to \phi(c, \ov{t})}
\]
Since $\IND(M')$ reduces to $\lambda c.\ M'\ c\ (\lambda b.\ \lambda x.\
\IND(M')\ b)$, by Lemma \ref{realredclosed} it suffices to show that for all $C, t$,
$M'\ t\ (\lambda b.\ \lambda x.\ \IND(M')\ b) \rrho \phi(C,
\ov{t})$. We proceed by induction on $\lambda$-rank of $C$. Take any $C, t$. 
By the inductive hypothesis, $M' \reals_\rho \forall c.\ (\forall b.\ b \ini c \to \phi(b, \ov{t}))
\to \phi(c, \ov{t})$, so $M' \downarrow \lambda c.\ N$ and $N[c:=t] \rrho \forall
b.\ b \ini C \to \phi(b, \ov{t})$. By Lemma \ref{realimpl}, it suffices to
show that $\lambda b.\ \lambda x.\ \IND(M')\ b \rrho \forall b.\ b \ini C \to \phi(b, \ov{t})$.
Take any $B, u$, $O \rrho B \ini C$, we need to show that
$\IND(M')[x:=O]\ u \rrho \phi(B, \ov{t})$. As $x \notin FV(M')$, it suffices
to show that $\IND(M')\ u \rrho
\phi(B, \ov{t})$, which, by Lemma \ref{realredclosed}, is equivalent to $M'\
u\ (\lambda b.\ \lambda x.\ \IND(M')\ b) \rrho \phi(B, \ov{t})$. 
As $O \rrho B \ini C$, the $\lambda$-rank of $B$ is less than the
$\lambda$-rank of $C$ and we get the claim by the inductive hypothesis.\qed
\end{enumerate}

\begin{corollary}[Normalization]\label{cornorm}
If $\proves M : \phi$, then $M \downarrow$. 
\end{corollary}

\begin{proof}
Take $\rho$ mapping all free propositional variables of $M$ to themselves
and all free first-order variables $a$ of $M$ to $(a, \emptyset)$. 
Then $\rho \models \p M : \phi$. By Theorem \ref{norm}, $\overline{M}[\rho]$
normalizes. By the definition of $\rho$, $\overline{M}[\rho] = \overline{M}$. By Lemma
\ref{erasurenorm}, $M$ normalizes.
\end{proof}

As the reduction system is deterministic, the distinction between strong and
weak normalization does not exist. If the reduction system is extended to
allow reductions anywhere inside the term, the Corollary \ref{cornorm}
shows only weak normalization. The counterexamples from \cite{jacsl2006}
adapted to $\lii$ show that \izfio\ does not strongly normalize and that non-well-founded
version does not normalize at all.

Our method of carrying the normalization proof is very different from the standard approach,
based on Girard's method of candidates \cite{GTL89}. As the candidates method is
usually used to show strong normalization of formal systems, it is unclear
if it could be applied to \izfio, given that it does not strongly normalize.
Although it might be possible to restate the realizability relation in terms
closer to the candidates method, we believe our account is easier to
understand and closer to its roots \cite{mccarty}. We will show how to apply
our method to show normalization of several weaker systems in the forthcoming \cite{jathesis}.

\ignore{\section{Applications}\label{secapp}}

The normalization theorem immediately provides the standard properties of
constructive set theories --- the disjunction property, the term existence
property, the set existence property and the numerical existence property.
Proofs are the same as in \cite{jacsl2006}; we only show the proofs of TEP
and SEP. 

\begin{corollary}[Term Existence Property]
If \izfio $\p \exists x.\ \phi(x)$, then there is a term $t$ such that \izfio $\p
\phi(t)$.
\end{corollary}
\begin{proof}
By the Curry-Howard isomorphism, there is a $\li$-term $M$ such that $\p M :
\exists x.\ \phi$. By Corollary \ref{corlz}, $M \downarrow v$ and $\p v :
\exists x.\ \phi$. By Canonical Forms, there is a pair $[t, N]$ such that
$\p N : \phi(t)$. Therefore, by the Curry-Howard isomorphism, \izfio $\p \phi(t)$.
\end{proof}

\begin{corollary}[Set Existence Property]
If \izfio $\p \exists x.\ \phi(x)$ and $\phi$ is term-free, then there is a term-free formula
$\psi(x)$ such that \izfio $\p \exists !x.\ \phi(x) \land \psi(x)$. 
\end{corollary}
\begin{proof}
By the previous corollary we have \izfio $\p \phi(t)$ for some term $t$.
Moreover, for any \izfio\ term $s$, there is a term-free defining formula
$\psi_s(x)$ such that \izfio $\p \psi_s(s) \land \exists !x.\ \psi_s(x)$.
Therefore \izfio $\p \exists !x.\ \phi(x) \land \psi_t(x)$. 
\end{proof}

In \cite{chol} we have shown how to use DP, NEP and TEP for the
purpose of program extraction. Thus our results establish \izfio\ as a valid
basis for a prover based on set theory with inaccessibles with the
capability of program extraction from constructive proofs.

\section{Related work}\label{others}

Several normalization results for impredicative constructive set theories much weaker than IZF exist. Bailin
\cite{bailin88} proved strong normalization of a constructive set theory
without the induction and replacement axioms. Miquel 
interpreted a theory of similar strength in a PTS (Pure Type System)
\cite{miquelpts}, where he also showed strong normalization of the calculus. This result was
later extended --- Dowek and Miquel \cite{dowek} interpreted a version of constructive
Zermelo set theory in a strongly normalizing \emph{deduction-modulo} system. 

In \cite{miquel}, Miquel interpreted \izfc\ without the $\in$-induction
axiom in a strongly-normalizing lambda calculus with types based on $F\omega.2$. 
It is unclear if Miquel's techniques could be used to prove any of DP, NEP,
SEP and TEP for the theory or to provide interpretations of ECC or CIC.

Krivine \cite{krivine} defined realizability using lambda calculus for classical set theory conservative
over ZF. The types for the calculus were defined. However, it seems to this
author that the types correspond to truth in the realizability model rather than to provable
statements in the theory. Moreover, the calculus does not even weakly normalize.

The standard metamathematical properties of theories related to IZF are well investigated.
Myhill \cite{myhill73} showed DP, NEP, SEP and TEP for IZF with Replacement and
non-recursive list of set terms. Friedman and \^S\^cedrov \cite{frsce1} showed SEP and
TEP for an extension of that theory with countable choice
axioms. Recently DP and NEP were shown for IZF with Collection
extended with various choice principles by Rathjen \cite{rathjenizf}.
However, the technique does not seem to be strong enough to provide TEP and SEP.

Powerful large set axioms (including the existence of class-many
inaccessibles) were added to IZF with Collection by Friedman and
\^S\^cedrov \cite{friedmanlarge}. The notion of an inaccessible set they
use differs from ours, as their inaccessibles must also model the
Collection axiom. We do not know if these two notions coincide. 
Both DP and NEP was shown for the resulting theories, but we do not think
that SEP and TEP could be proved with their technique.

Inaccessible sets were also investigated in the context of weaker,
predicative CZF (Constructive Zermelo-Fraenkel). Crosilla and Rathjen
\cite{crosilla02} showed that the power of inaccessible
set axioms might be closely linked to the $\in$-induction axiom. They 
proved that inaccessible sets added to CZF with $\in$-induction taken away
do not add any proof-theoretical power. 

\section*{Acknowledgements}

I would like to thank my advisor, Bob Constable, for comments and support,
Richard Shore for helpful discussions, David Martin for commenting on
the early stages of this research and anonymous referees for their comments.

\bibliographystyle{alpha}
\bibliography{rc.bib}

\end{document}